\documentclass[10pt,letterpaper]{article}
\usepackage[top=0.85in,left=1.55in,right=0.35in, footskip=0.75in]{geometry}
\usepackage{amsmath,amssymb}

\usepackage{changepage}

\usepackage[utf8]{inputenc}

\usepackage{textcomp,marvosym}

\usepackage{nameref,hyperref}

\usepackage[right]{lineno}

\usepackage{microtype}
\DisableLigatures[f]{encoding = *, family = * }

\usepackage[table]{xcolor}

\usepackage{array}

\newcolumntype{+}{!{\vrule width 2pt}}

\newlength\savedwidth

\usepackage[aboveskip=1pt,labelfont=bf,labelsep=period,singlelinecheck=off]{caption}

\usepackage[sorting=none,url=false,isbn=false,doi=false,eprint=false]{biblatex}
\bibliography{bib/biblio}
\AtEveryBibitem{
 \clearlist{language}
 \clearlist{editor}
}

\makeatletter
\renewcommand{\@biblabel}[1]{\quad#1.}
\makeatother

\usepackage{lastpage,fancyhdr,graphicx}
\usepackage{epstopdf}
\fancyheadoffset[L]{1.25in}
\fancyfootoffset[L]{1.25in}

\usepackage[section]{placeins}
\usepackage{float}
\usepackage{booktabs}
\usepackage{bookmark}
\usepackage{multicol}

\usepackage{url}
\usepackage{breakurl}

\allowdisplaybreaks

\begin{document}

\vspace*{0.2in}

\begin{flushleft}
{\Large
\textbf\newline{Local dendritic balance enables learning of efficient representations in networks of spiking neurons} 
}
\newline
\\
Fabian A. Mikulasch\textsuperscript{1*},
Lucas Rudelt\textsuperscript{1*},
Viola Priesemann\textsuperscript{1,2\textdagger}
\\

\bigskip
\textbf{1} Max-Planck-Institute for Dynamics and Self-Organization, Göttingen, Germany
\\
\textbf{2} Bernstein Center for Computational Neuroscience (BCCN) Göttingen
\\
\bigskip

* These authors contributed equally \\
\textdagger{} Corresponding author: viola.priesemann@ds.mpg.de

\end{flushleft}

\hypertarget{abstract}{%
\section*{Abstract}\label{abstract}}

How can neural networks learn to efficiently represent complex and high-dimensional inputs via local plasticity mechanisms? Classical models of representation learning assume that input weights are learned via pairwise Hebbian-like plasticity. Here, we show that pairwise Hebbian-like plasticity only works under unrealistic requirements on neural dynamics and input statistics.
To overcome these limitations, we derive from first principles a learning scheme based on voltage-dependent synaptic plasticity rules.
Here, inhibition learns to locally balance excitatory input in individual dendritic compartments, and thereby can modulate excitatory synaptic plasticity to learn efficient representations.
We demonstrate in simulations that this learning scheme works robustly even for complex, high-dimensional and correlated inputs, and with inhibitory transmission delays, where Hebbian-like plasticity fails.
Our results draw a direct connection between dendritic excitatory-inhibitory balance and voltage-dependent synaptic plasticity as observed \textit{in vivo}, and suggest that both are crucial for representation learning.

\hypertarget{significance}{%
\section*{Significance}\label{significance}}
Neurons have to represent an enormous amount of sensory information. In order to represent this information efficiently, neurons have to adapt their connections to the sensory inputs. 
An unresolved problem is how this learning is possible when neurons fire in a correlated way. Yet, these correlations are ubiquitous in neural spiking, either because sensory input shows correlations, or because perfect decorrelation of neural spiking through inhibition fails due to physiological transmission delays. We derived from first principles that neurons can, nonetheless, learn efficient representations if inhibition modulates synaptic plasticity in individual dendritic compartments.
Our work questions pairwise Hebbian plasticity as a paradigm for representation learning, and draws a novel link between representation learning and a dendritic balance of input currents.

\newpage

\hypertarget{introduction}{%
\section*{Introduction}\label{introduction}}
\pdfbookmark[section]{Introduction}{introduction}
Many neural systems have to encode high-dimensional and complex input signals in their
activity. It has long been hypothesized that these encodings
are highly efficient, that is, neural activity faithfully represents the input
while also obeying energy and information constrains
\autocite{rosenblith_possible_2012}. For populations of spiking neurons,
such an efficient code requires two central features: first, neural
activity in the population has to be coordinated, such that no spike is
fired superfluously \autocite{sengupta_balanced_2013}; second,
individual neural activity should represent elementary features in the
sensory input signal, which reflect the statistics of the stimuli
\autocites{atick1990towards}{olshausen_emergence_1996}. 
How can this
coordination and these efficient representations emerge through local
plasticity rules?

To coordinate neural spiking, a population has to provide
information about the population response locally at each neuron. More specifically, for an
efficient encoding, input signals should not be encoded redundantly by
different neurons, which in many cases means the population should decorrelate
responses, e.g. through lateral inhibition. This decorrelation can be realized
by networks with excitatory-inhibitory (E-I) balance
\autocite{vreeswijk1998chaotic}. Recently it became clear that spiking
networks can find an especially cooperative encoding by learning a
tight E-I balance \autocite{deneve_efficient_2016}. This gave rise to a
novel perspective on inhibitory plasticity
\autocite{vogels_inhibitory_2011} and the E-I balance in biological
networks \autocites{dehghani_dynamic_2016}{haider_neocortical_2006}: By
learning E-I balance, neural responses to input signals are distributed
over the population, which, from an information-theoretic and
physiological perspective, renders the code efficient
\autocite{sengupta_balanced_2013}.

To efficiently encode high-dimensional input signals, it is additionally
important that neural representations are adapted to the statistics of
the input. Early studies of recurrent networks showed that such representations 
can be found through Hebbian-like learning of input weights
\autocites{foldiak1990forming}{linsker1992local}. With Hebbian learning
the repeated occurrence of patterns in the input is associated with
postsynaptic activity, causing neurons in the population to become
detectors of recurrent elementary features. However, this learning also requires the
decorrelation of responses through inhibition. Inhibition
\emph{indirectly} guides the learning process by forcing neurons to
fire for distinct features in the input. Recent efforts rigorously
formalized this idea for models of spiking neurons in balanced networks
\autocite{brendel2020learning} and spiking neurons sampling from
generative models
\autocites{nessler_bayesian_2013}{bill_distributed_2015}{nessler2009stdp}{kappel_stdp_2014}.
The great strength of these approaches is that the learning rules
can be derived from first principles, and turn out to be similar to spike timing
dependent plasticity (STDP) curves that have been measured experimentally.

However, to enable the learning of efficient representations, these models have
strict requirements on network dynamics and input. Most crucially,
inhibition has to ensure that neural responses are sufficiently
decorrelated. In the neural sampling approaches, learning therefore
relies on strong winner-take-all dynamics. In the framework of balanced
networks, inhibition has to be nearly instantaneous and elementary features
in the input have to be uncorrelated. These requirements are likely not
met in realistic situations.

We here propose a mechanism that overcomes these limitations
and enables spiking networks to learn efficient representations. We
suggest that inhibition can \emph{directly} guide the learning of input
weights by learning to locally balance specific inputs on dendrites. The
resulting balanced dendritic potentials can be used to incorporate
information about the population code into the learning of single input
weights. In simulations of spiking neural networks we demonstrate the benefits
of this learning scheme over Hebbian-like learning for the efficient
encoding of high-dimensional inputs. Hence, we extend the idea that
balanced state inhibition provides information about the population code
locally, and show that not only can it be used to distribute neural
responses over a population, but also for an improved learning of input
weights.

\newpage

\hypertarget{results}{%
\section*{Results}\label{results}}
\pdfbookmark[section]{Results}{results}
\begin{figure}[t]
\hypertarget{fig:intro}{%
\centering
\includegraphics[scale=0.7]{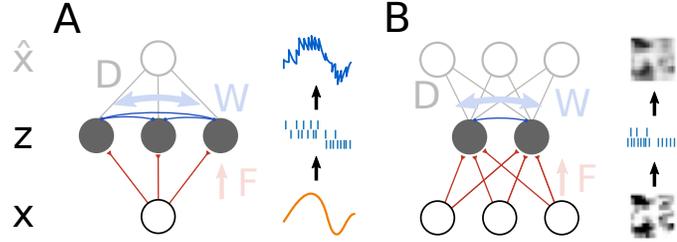}
\caption{\emph{For representation learning, the task is to efficiently encode high dimensional, analog input signals with spikes.}
\textbf{A} Low dimensional, analog input signals $\mathbf{x}$ can be efficiently encoded by the responses of a large population of spiking neurons $\mathbf{z}$. To that end, neurons couple to the input via weights $F$ and inhibit each other via weights $W$. From the encoding an external observer can decode an approximation
$\mathbf{\hat x}$ of the original input signal $\mathbf{x}$ via a linear transformation $D$.
\textbf{B} In this paper, we ask how high dimensional, analog
input signals $\mathbf{x}$ can be efficiently encoded by a small population of spiking
neurons. For this more complex task the neurons not only
have to translate the continuous input signal into a spiking representation,
they also have to compress it by adapting their input weights $F$ to the input statistics.}\label{fig:intro}
}
\end{figure}

The goal in this paper is to efficiently encode a continuous
high-dimensional input signal by neural spiking. In the following, we
will explain how neurons can learn efficient representations of these
inputs through local plasticity mechanisms. We will first show how
E-I balance can guide neural spiking in order to distribute the
encoding over the population. We will then show how E-I balance
on the level of dendrites can guide the learning of efficient
representations in the input weights.

\hypertarget{background-efficient-encoding-by-spiking-neurons-through-tight-E-I-balance}{%
\subsection*{Background: Efficient encoding by spiking neurons with tight E-I balance}\label{background-efficient-encoding-by-spiking-neurons-through-tight-E-I-balance}}
\pdfbookmark[subsection]{Background: Efficient encoding by spiking neurons with tight E-I balance}{background-efficient-encoding-by-spiking-neurons-through-tight-E-I-balance}

\hypertarget{setup}{%
\subsubsection*{Setup}\label{setup}}
\pdfbookmark[subsubsection]{Setup}{setup}

Continuous inputs $\mathbf{x}(t)$ drive a recurrently connected
spiking neural network which encodes the inputs through responses $\mathbf{z}(t)$ (Fig~\ref{fig:intro}A). Input weights $F_{ji}$ indicate how strongly excitatory input $x_i(t)$ couples to neuron $j$ and lateral inhibitory weights $W_{jk}$ provide negative coupling between the neurons. Neurons in the network encode the inputs by emitting spikes, which then elicit postsynaptic potentials (PSPs) $\mathbf{z}(t)$. The PSPs are modeled as a sum of exponentially decaying depolarizations $z_j(t)=\sum_{t_s^j \leq t - \Delta t} \exp(-\frac{t-\Delta t-t_s^j}{\tau})$ with decay time $\tau$ for each spike of neuron $j$ at times $t_s^j$. PSPs arrive after one timestep $\Delta t$, which we interpret as a finite transmission delay of neural communication. Our model is similar to those in previous studies of balanced spiking networks \autocites{brendel2020learning}{boerlin2013predictive}. 

To test whether the input is well encoded, we consider the best linear
readout $\mathbf{\hat x}(t)=D\mathbf{z}(t)$ of inputs from the neural response and quantify the
mean decoder loss
\begin{equation}\begin{aligned}
\mathcal{L}= \frac{1}{N_x} \left\langle||\mathbf{x}(t)-\mathbf{\hat x}(t)||^2\right\rangle_t= \frac{1}{N_x} \left\langle ||\mathbf{x}(t)-D\mathbf{z}(t)||^2\right\rangle_t
\end{aligned},\label{eq:decoderloss}\end{equation}
where $N_x$ is the number of inputs. 
It is important to note that the
readout is not part of the network, but only serves as a guidance 
to define a computational goal that can
be reached autonomously via local plasticity rules. Hence learning an efficient code
amounts to minimizing $\mathcal{L}$ via local plasticity rules on
$F_{ji}$ and $W_{jk}$ given the best decoder $D$.

\hypertarget{spiking-neuron-model}{%
\subsubsection*{Spiking neuron model}\label{spiking-neuron-model}}
\pdfbookmark[subsubsection]{Spiking neuron model}{spiking-neuron-model}

Spiking neurons are modeled as stochastic leaky integrate-and-fire
neurons. Stochastic spiking is important to
increase robustness during learning, and allows a direct link to neural
sampling and unsupervised learning via expectation-maximization (see SI). A neuron
$j$ emits spikes stochastically with a probability that depends on its
membrane potential $u_j(t)$ according to

\begin{equation}p_{spike}(u_j(t)) \propto \exp\left(\frac{u_j(t)-T_j}{\Delta u}\right).\label{eq:stochasticspike}\end{equation}

\begin{figure}
\hypertarget{fig:learningrules}{%
\centering
\includegraphics[scale=0.65]{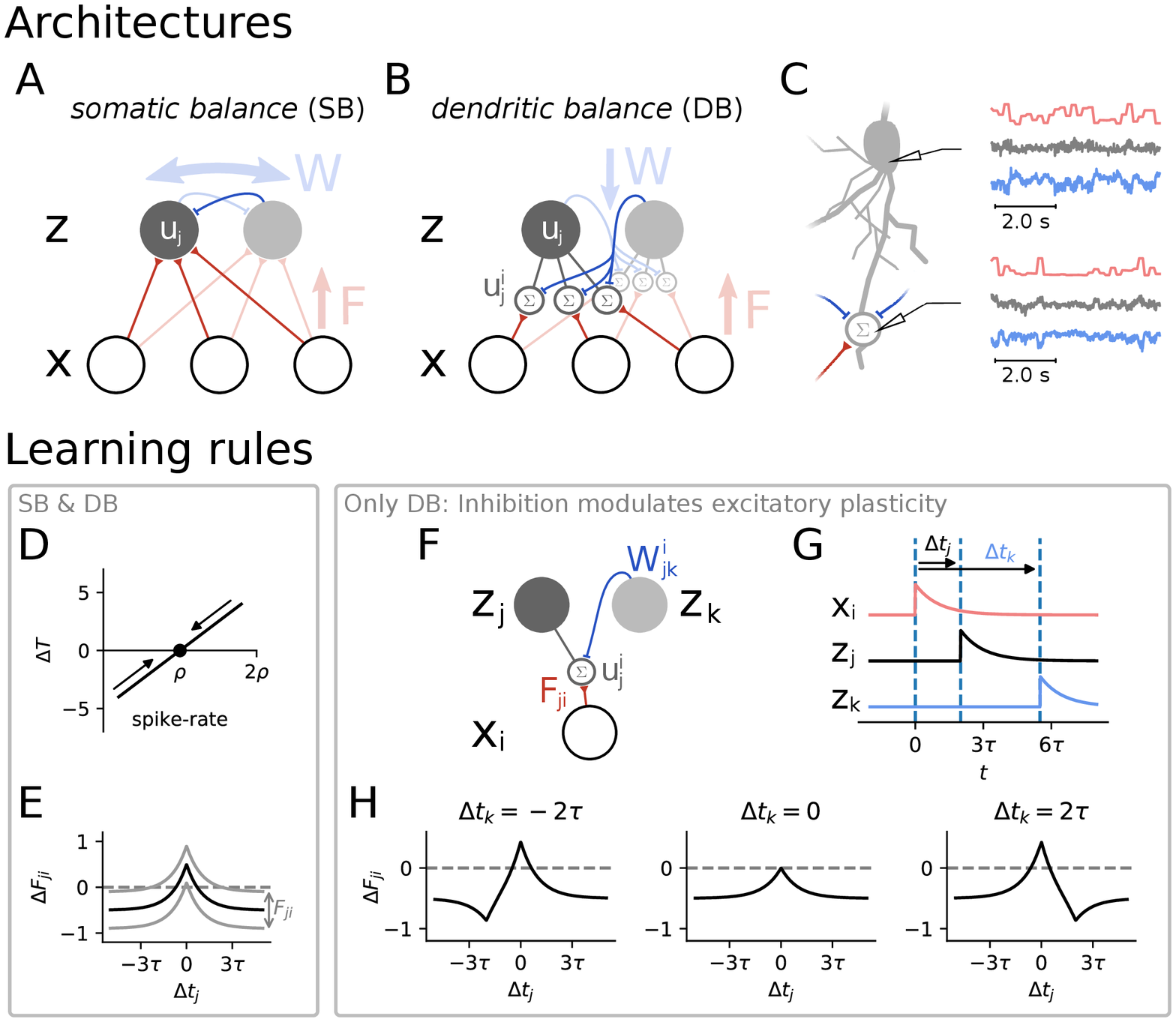}
\caption{\emph{We compare learning in two network models, a classical model of point neurons with somatic balance, and a model with dendritic balance.}
\textbf{A} In the classical model (SB), neurons (grey circles) with outputs $\mathbf{z}$ receive feed-forward network inputs $\mathbf{x}$ (white circles) and are coupled via lateral inhibition. Inhibitory weights $W$ are adapted to balance excitatory currents at the soma with membrane potential $u_j$, which ensures an efficient spike encoding.
\textbf{B} In our proposed model (DB), neurons receive inputs at specific dendritic compartments. Inhibition learns to balance input currents locally at the dendrites. 
This leads to dendritic potentials $u^i_j$ that are proportional to the coding error for specific inputs, and therefore can be used to learn input weights.
\textbf{C} After learning, local excitatory (red) and inhibitory (blue) currents have adapted to tightly balance each other at individual dendrites (bottom). This dendritic balance also results in a somatic balance of inputs (top). Here we show a neuron from a network with 80 neurons coding for natural images.
\textbf{D} In both models a rapid compensatory mechanism ensures that every neuron fires with rate $\rho$. If any neuron spikes too rarely, its threshold $T_j$ is lowered; if it spikes too often $T_j$ is increased. This mechanism allows that the performances of SB and DB can be compared, but is not required for stable network function (see Methods).
\textbf{E} For learning input weights in the classical model a Hebbian-like STDP rule increases or decreases weights $F_{ji}$ depending on the time difference between pre- and postsynaptic spikes $\Delta t_j$, and the weight $F_{ji}$ itself. If $F_{ji}$ is high or low, this shifts plasticity towards depression or potentiation, respectively. The same learning rule applies to the proposed model, if a neuron does not simultaneously receive any inhibition.
\textbf{F-H} Illustration of how inhibition modulates input plasticity in the proposed model for a network of two coding neurons $z_j$ (with one dendritic compartment) and $z_k$ and one input neuron $x_i$. 
\textbf{F} The excitatory weight $F_{ij}$, as well as the inhibitory weight $W_{jk}^i$, attach to the same dendritic potential $u^i_j$. 
\textbf{G} We consider the following example where three spikes are fired: $x_i$ at $t=0$, $z_j$ at $t=\Delta t_j$ and $z_k$ at $t=\Delta t_k$. 
\textbf{H} The total change in the weight $F_{ji}$ depends not only on the spike-time difference $\Delta t_j$ between the input and the postsynaptic neuron, but also on the relative inhibitory spike-time $\Delta t_k$. 
In general, if $z_j$ and $z_k$ spike close together, $F_{ji}$ will tend to be depressed. All
weight-changes were calculated with $F_{ji} = -W_{jk}^i = 0.5$.}\label{fig:learningrules}
}
\end{figure}

When the membrane potential approaches the firing threshold $T_j$, the
firing probability increases exponentially, where $\Delta u$ regulates the
stochasticity of spiking. For increasing $\Delta u$ the spike emission
becomes increasingly random, whereas for $\Delta u \rightarrow 0$ one
recovers the standard leaky integrate-and-fire (LIF) neuron with sharp threshold. The membrane potential itself is modeled as a linear sum
of the feed-forward inputs $x_i(t)$ and lateral inhibition $z_k(t)$,
i.e.
\begin{equation}u_j(t)=\underbrace{\textstyle{\sum_i} F_{ji}x_i(t)}_{\text{excitatory input}} + \underbrace{\textstyle{\sum_{k}} W_{jk} z_k(t)}_{\text{lateral inhibition}}.\label{eq:somaticpotential}\end{equation}
In order to fix the number of spikes for an efficient code, the
average firing rate of each neuron was controlled via 
the threshold $T_j$ (Fig~\ref{fig:learningrules}C). 

\hypertarget{somatic-balance-enables-an-efficient-encoding}{%
\subsubsection*{Somatic balance enables an efficient
encoding for low-dimensional input  signals}\label{somatic-balance-enables-an-efficient-encoding}}
\pdfbookmark[subsubsection]{Somatic balance enables an efficient
encoding for low-dimensional input  signals}{somatic-balance-enables-an-efficient-encoding}

Spiking neurons can implement an efficient encoding for low dimensional input signals by tightly balancing feed-forward inputs and lateral inhibition at the soma \autocite{deneve_efficient_2016}. In fact, for one dimensional inputs a tight balance directly implies an efficient encoding. 
In this case, the inhibitory weights can be rewritten as $W_{jk} = - F_{j} D_{k}$ for fixed input weights $F$ and some decoder $D$. Similarly, the membrane potentials can be written as $u_j(t) = F_{j} \left(x(t)-\hat{x}(t)\right)$, with $\hat{x}(t) = \sum_k D_{k} z_k(t)$.
Thus, when the membrane potentials are tightly balanced, i.e. $u_j(t)\approx 0$ most of the time, then the two requirements of an efficient code are fulfilled: First, the decoder $D$ is able to accurately reconstruct the input from the network activity. Second, since the neurons are only driven by the coding error $x(t)-\hat{x}(t)$, no superfluous spike is fired that does not improve the encoding.

This motivates that networks can autonomously find an efficient encoding by enforcing a tight balance through local inhibitory plasticity (see SI for derivation) \begin{equation}\Delta W_{jk} \propto -z_k u_j =-z_k\left( \textstyle{\sum_i} F_{ji}x_i + \textstyle{\sum_{l}} W_{jl} z_l\right) \quad \text{(somatic balance)}.
\label{eq:somaticbalance}\end{equation}
Hence, when neuron $k$ is active and the somatic potential of neuron $j$ is out of balance, i.e. $u_j(t)\neq 0$, the weight $W_{jk}$ changes to balance $u_j(t)$. Note, that all neurons have an autapse that learns to balance their own membrane potentials, which can alternatively be interpreted as an approximate membrane potential reset after spiking.
As was shown in \autocite{brendel2020learning},
tight E-I balance is sufficient to efficiently encode low
dimensional input signals (Fig~\ref{fig:intro}A).

\hypertarget{main-result-efficient-representations-by-spiking-neurons-through-plastic-input-synapses}{%
\subsection*{Learning efficient representations with plastic input synapses}\label{main-result-efficient-representations-by-spiking-neurons-through-plastic-input-synapses}}
\pdfbookmark[subsection]{Learning efficient representations with plastic input synapses}{main-result-efficient-representations-by-spiking-neurons-through-plastic-input-synapses}

While the exact choice of input weights $F$ is irrelevant for low
dimensional inputs, it becomes
increasingly important for high dimensional inputs with complex statistical structure (Fig~\ref{fig:intro}B). 
It is possible to derive a plasticity rule for input
synapses $F_{ji}$ that minimizes the decoder loss $\mathcal{L}$ via
gradient descend, which yields
\begin{equation}\Delta F_{ji}\propto z_j(x_i-\hat x_i) =  z_j(x_i-\textstyle{\sum_k} D_{ik}z_k) .\label{eq:lr}\end{equation}
Intuitively, this rule drives neuron $j$ to correlate its output
$z_j$ to input $x_i$, except if the population is already encoding
it. To extract the latter information, the plasticity rule requires a
decoding $\hat{x}_i=\sum_k D_{ik}z_k$, which contains information
about the neural code for input $i$ of all other neurons in the population. 

We thus conclude that an efficient code relies on information about
other neurons in two ways: (i)~neurons need to know what is already
encoded to avoid redundancy in spiking (dynamics), (ii)~plasticity of
input connections requires to know what neurons encode about specific
inputs to avoid redundancy in representation (learning). While
inhibitory weights $W_{jk}$ for efficient spiking dynamics (i) can be
learned locally (Eq \ref{eq:somaticbalance}), learning input synapses
$F_{ji}$ (ii) is not feasible locally for point neurons, since they lack knowledge about the population code for single inputs $x_i$.

In the following, we will introduce the main result of this paper:
similarly to efficient spiking through a tight balance of \emph{all}
excitatory and inhibitory inputs at the soma, local learning of
efficient representations can be realized by tightly balancing
\emph{specific} excitatory inputs with lateral inhibition.
Physiologically, we argue that this corresponds to spatially separated
inputs at different dendritic compartments, where dendritic inhibition
locally balances the membrane potential. We contrast this local
implementation of the correct gradient of the decoder loss with a common
local approximation of the gradient, which is necessary for neurons with
somatic balance only.

\hypertarget{somatic-balance-alone-requires-a-local-approximation}{%
\subsubsection*{Somatic balance alone requires an
approximation for local learning}\label{somatic-balance-alone-requires-a-local-approximation}}
\pdfbookmark[subsubsection]{Somatic balance alone requires an
approximation for local learning}{somatic-balance-alone-requires-a-local-approximation}

Since somatic balance alone cannot provide information about other
neurons at the input synapses, previous approaches used a local
approximation to $\Delta F_{ji}$ where only pre- and postsynaptic
currents are taken into account (Fig~\ref{fig:learningrules}E)
\begin{equation}\begin{aligned}
 \Delta F_{ji}\propto z_j(x_i-F_{ji}z_j) \quad(\text{Hebbian-like learning}).
\end{aligned}\label{eq:localapproximation}\end{equation}
We will refer to this setup as somatic balance (SB), because
inhibition is mediated by inhibitory connections that balance the
somatic potential $u_j(t)$.

The above learning rule is exact when simultaneous coding, and thus
non-local dependencies during learning, are not present. This is the
case when only a single PSP $z_j(t)$ is nonzero at a time, e.g.~in
winner-take-all circuits with extremely strong inhibition
\autocite{bill_distributed_2015}, or when the PSP is extremely short
\autocite{nessler_bayesian_2013}. The learning rule becomes also
approximately exact when neural PSPs $\mathbf{z}(t)$ in the encoding
are uncorrelated \autocites{foldiak1990forming}{brendel2020learning}.
However, these are strong demands on the dynamics of the network which
ultimately limit its coding versatility.

\hypertarget{dendritic-balance-allows-local-learning-of-efficient-representations}{%
\subsubsection*{Dendritic balance allows local learning of efficient
representations}\label{dendritic-balance-allows-local-learning-of-efficient-representations}}
\pdfbookmark[subsubsection]{Dendritic balance allows local learning of efficient representations}{dendritic-balance-allows-local-learning-of-efficient-representations}

When neural PSPs $\mathbf{z}(t)$ in the population are correlated,
learning efficient representations at input synapses requires that
information about the population code for this input is available at the synapse. To this
end, we introduce local dendritic potentials $u_j^i$ at synapses
$F_{ji}$, and couple neurons $k$ via dendritic inhibitory
connections $W_{jk}^i$ to these membrane potentials (Fig
\ref{fig:learningrules}B). The somatic membrane potential is then
realized as the linear sum of the local dendritic potentials
\begin{equation}\begin{gathered}
u_j(t) = \textstyle{\sum_i} u_j^i(t) \\
u^i_j(t) = \underbrace{F_{ji} x_i(t)}_{\text{excitatory input}} + \underbrace{\textstyle{\sum_k} W^i_{jk} z_k(t)}_{\text{dendritic inhibition}}.
\end{gathered}\label{eq:dendriticpotential}\end{equation}
Note
that this amounts only to a refactoring of the equation for the somatic
membrane potential and does not change the computational power of the neuron.
Given a network using dendritic inhibition with inhibitory
weights $W^i_{jk}$, a network using somatic inhibition with weights
$W_{jk}=\sum_i W^i_{jk}$ is equivalent. Hence, any improvement in the
neural code through dendritic balance is due to an improvement in the
learning of feed-forward weights.

The local integration of dendritic inhibition allows to use the same
trick as before: by enforcing a tight E-I balance locally, dendritic inhibition will try to cancel the
input as well as possible. Thereby, dendritic inhibitory weights
$W_{jk}^i$ will automatically learn the best possible decoding of the
population activity $\mathbf z$ to the input $F_{ji}x_i$. This leads to a
local potential that is proportional to the coding error
$u^i_j = F_{ji} (x_i - \hat x_i)$. In terms of synaptic inhibitory plasticity this is
realized by
\begin{equation} \Delta W^i_{jk} \propto -z_k u^i_j \quad \text{(dendritic balance)}.\label{eq:dedriticbalance}\end{equation}
Thus, the
dendritic membrane potential $u^i_j$ can be used to find the correct gradient $\Delta F_{ji}$ from Eq
\ref{eq:lr} locally
\begin{equation}\Delta F_{ji} \propto \frac{1}{F_{ji}} z_j u^i_j \quad\text{(learning by errors)}.\label{eq:correctlearning}\end{equation}

As can be seen, the learning rules for input and inhibitory weights both
rely on the local dendritic potential, which they also influence. This
enables local inhibition to modulate feed-forward plasticity. However,
in our model this also requires the cooperation of the excitatory and
inhibitory weights during learning. We propose two different
implementations which ensure this cooperation, by learning inhibitory
weights on a faster, or on the same timescale as excitatory weights (see SI). We show that these two approaches yield similar results which
equal the analytical solution (Eq \ref{eq:lr}) in performance (see Fig~\ref{fig:a1},\ref{fig:a2}).

It is possible to integrate the learning rules which depend on membrane
potentials over time and obtain learning rules which depend on the
relative spike timings of multiple neurons. If we only consider one
input neuron and one coding neuron, learning with dendritic balance
and somatic balance yield the same spike timing dependent plasticity
rule. This rule is purely symmetric and strengthens the connection when
both neurons fire close in time (Fig~\ref{fig:learningrules}E).
However, if the spike of the excitatory input neuron is accompanied
by an inhibitory spike in the coding population, the spike timing
dependent rule breaks symmetry (Fig~\ref{fig:learningrules}H). This
shows how learning with dendritic balance can take more than pair-wise
interactions into account when learning weights to enable the neuron to
find its place in the ongoing activity.

\hypertarget{Simulation-experiments}{%
\subsection*{Simulation experiments}\label{Simulation experiments}}
\pdfbookmark[subsection]{Simulation experiments}{Simulation-experiments}

To illustrate the differences that arise between the networks using
somatic balance (SB) and dendritic balance (DB) during learning, we set
up several coding tasks of increasing complexity. In order to facilitate
the interpretation of the input features learned by the neurons, we
tasked the neural population to code for images. The images were
presented as constant input signals over time and faded in between
presentations to avoid discontinuities. Learning was
performed on-line in an unsupervised fashion and single neurons
consistently learned to represent elementary features of the input stimuli.
We quantified performance by measuring the decoder loss of this neural
code on a separate set of test stimuli with plasticity rules turned off.

\hypertarget{simple-stimuli-are-encoded-equally-well-by-networks-using-somatic-or-dendritic-balance}{%
\subsubsection*{Simple stimuli are encoded equally well by networks using
somatic or dendritic
balance}\label{simple-stimuli-are-encoded-equally-well-by-networks-using-somatic-or-dendritic-balance}}
\pdfbookmark[subsubsection]{Simple stimuli are encoded equally well by networks using somatic or dendritic balance}{simple-stimuli-are-encoded-equally-well-by-networks-using-somatic-or-dendritic-balance}

\begin{figure}
\hypertarget{fig:mnist}{%
\centering
\includegraphics[scale=0.5]{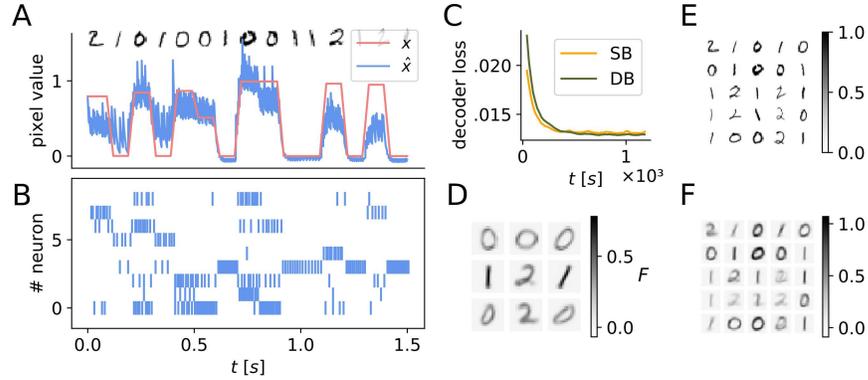}
\caption{\emph{Simple stimuli are encoded equally well by networks using
somatic or dendritic balance.} In this simulation experiment, networks consisting of
9 coding neurons encoded 16x16 images of digits 0, 1 and 2 from the MNIST dataset. 
\textbf{A} Input signal $x_i$ and decoded signal $\hat{x}_i$ for a single pixel
$i$ in the center of the image. MNIST digits were presented as constant input signals for $70\,\text{ms}$ and faded for $30\,\text{ms}$ to avoid discontinuities. The decoded signal tracks the input reasonably well given the very limited capacity of the the network. 
\textbf{B} Spike train of all neurons in the network in response to the input signal in A.
\textbf{C} Decoder loss decreases with neural plasticity for both models using either
somatic balance (SB), or dendritic balance (DB). Both models have very similar learning speed and final performance. The decoder loss was calculated on a fixed
set of test images.
In this example, DB is realized by simultaneous learning of input and inhibitory weights (see SI).
\textbf{D} Input weights of the 9 coding neurons in the DB network after learning. Every neuron becomes specific for a certain prototypical digit.
\textbf{E} Sample of input images from the MNIST dataset.
\textbf{F} Reconstructions $\mathbf{\hat x}$ of the input images shown
in E. The reconstructions presented here are calculated by
averaging the decoded signal during input signal presentation over $70\,\text{ms}$.
In A,B,D and F we show results of the converged DB network.
}\label{fig:mnist}
}
\end{figure}

In a first test we performed a comparison on the MNIST dataset of
handwritten digits (Fig~\ref{fig:mnist}E). We restricted the dataset
to the digits 0, 1 and 2, which were encoded by 9 coding neurons. After
learning, the input weights of both networks had converged to detect
prototypical digits (Fig~\ref{fig:mnist}D) and the codes and coding
performances were approximately equal (Fig~\ref{fig:mnist}C).
Since images were rarely encoded by more than one or two neurons (Fig~\ref{fig:mnist}A, B), interactions in the population were small and thus
the learning rules found similar solutions.

\hypertarget{dendritic-balance-can-disentangle-weak-correlations}{%
\subsubsection*{Dendritic balance can disentangle correlated features}\label{dendritic-balance-can-disentangle-weak-correlations}}
\pdfbookmark[subsubsection]{Dendritic balance can disentangle correlated features}{dendritic-balance-can-disentangle-weak-correlations}

\begin{figure}
\hypertarget{fig:bars}{%
\centering
\includegraphics[scale=0.6]{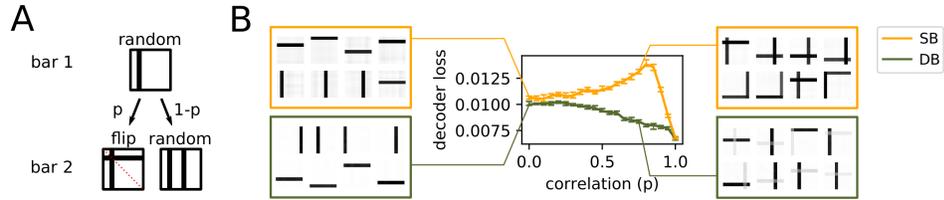}
\caption{\emph{Dendritic balance enables neurons to disentangle feature correlations in the input.} In this simulation experiment, 16 neurons code for 8x8 images containing two random out of 16 possible bars. 
Thus, optimally, every neuron codes for a single bar. 
\textbf{A} Creation of input images with correlation between features. Two bars are selected in succession and added to the image. With probability $p$ the bars are symmetric around the top-left to bottom-right diagonal axis. With probability $1-p$ the two bars are chosen randomly. 
\textbf{B} Decoder loss after learning for different correlation strengths. Displayed is the median decoder loss for 50 different realizations for each data point; error-bars denote 95\% bootstrap confidence intervals. 
On the sides, 8 out of all 16 converged input weights are shown for representative
networks. When correlations between bars are present, the representations learned by SB overlap, while DB still learns efficient single bar representations.}\label{fig:bars}
}
\end{figure}

Our theoretical results suggest that DB networks should find a better
encoding than SB networks when weak correlations between elementary
features are present in the stimuli. To test this, we devised a
variation of Földiak's bar task \autocite{foldiak1990forming}, which is
a classic independent component separation task. In the original task
neurons encode images of independently occurring but overlapping
vertical and horizontal bars. Since the number of neurons is equal to
the number of possible bars in the images, each neuron should learn to
represent a single bar to enable a good encoding. We kept this basic
setup but additionally we introduced between-bar correlations for
selected pairs of bars (Fig~\ref{fig:bars}A). We then could vary the
correlation strength $p$ between the bars within the pairs to render
them easier or harder to separate.

The simulation results indeed showed that the performance of SB, but not
of DB, deteriorates when elementary features are correlated (Fig
\ref{fig:bars}B). The decoder loss for SB grows for increasing $p$ and
reaches its maximum at about $p=0.8$. This is because Hebbian-like
learning (as used in SB) correlates a neuron's activity with the
appearance of patterns in the input signal, irrespective of the population
activity. The correlation between two bars therefore can lead a neuron
which initially is coding for only one of the bars to incorporate also
the second bar into its receptive field. Thus, with a certain
correlation strength $p$ between bars the receptive fields of neurons
start to collapse. For $p>0.8$ the decoder loss decreases, as here the
occurrence of specific pairs of bars becomes so likely, that the
collapsed representations reflect the statistics of the images again. In
contrast, DB enables neurons to communicate which part of the input signal
they encode and hence they consistently learn to code for single bars.
Accordingly, the decoder loss for DB is smaller than for SB for every
correlation strength of bars.

\hypertarget{dendritic-balance-improves-learning-for-images-of-natural-scenes}{%
\subsubsection*{Dendritic balance improves learning for images of natural
scenes}\label{dendritic-balance-improves-learning-for-images-of-natural-scenes}}
\pdfbookmark[subsubsection]{Dendritic balance improves learning for images of natural scenes}{dendritic-balance-improves-learning-for-images-of-natural-scenes}

\begin{figure}
\hypertarget{fig:scenes}{%
\centering
\includegraphics[scale=0.6]{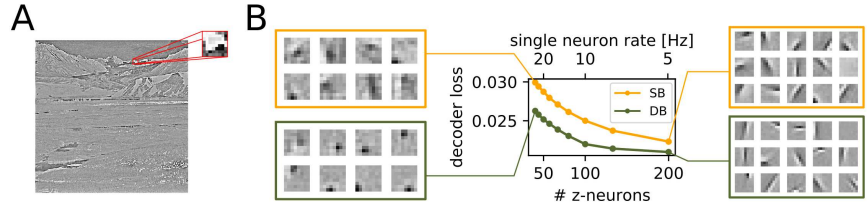}
\caption{
\emph{Dendritic balance enables learning of efficient representations of complex, natural stimuli.} \textbf{A} Exemplary image from the natural-scenes data-set.
We extracted 16x16 pixel images from a set of whitened pictures of
natural scenes \autocite{olshausen_emergence_1996}, scaled them down to 8x8
pixels and applied a non-linearity (see SI). 
\textbf{B} Decoder loss after learning of SB and DB networks featuring varying numbers of coding neurons, while keeping the population rate constant at $1000\,\text{Hz}$. 
On the sides we show exemplary converged input weights. For a large number
of coding neurons (right) both learning schemes yield similar representations, but performance is slightly better for DB. 
For a small number of neurons (left) DB learns more refined representations with substantially reduced decoder loss as compared to SB. The reason is that for a small number of neurons the learned features are more correlated and consequently are harder to disentangle. 
Notably, different amounts of neurons result in different coding strategies.
}\label{fig:scenes}  
}
\end{figure}

We expected to see a similar difference between SB and DB networks when
complex stimuli are to be encoded. In a third experiment we therefore
tested the performance of the networks encoding images of natural
scenes. We used 8x8 pixel images cut out from a set of pictures of
landscapes and vegetation with suitable preprocessing (Fig
\ref{fig:scenes}A). To also test whether the amount of compression
(number of inputs vs.~number of coding neurons) would affect SB and DB
networks differently, we varied the number of coding neurons while
keeping the population rate fixed at 1000Hz. This way, only the
compression, and not also the total number of spikes, has an effect on the performance of the networks.

The simulations showed that complex stimuli can be represented better by
DB networks compared to SB networks. This
difference becomes larger the higher the compression of the input signal by
the network is (Fig~\ref{fig:scenes}B). This effect seems to be
related to the observations we made in the bar task: Networks with few
coding neurons have to learn correlated features, which renders SB less
appropriate. We found that SB networks consistently needed about twice as many neurons
to achieve a similar coding performance as DB networks.

We also note that the amount of compression affects the strategy neurons
resort to in order to encode the images. When the number of coding
neurons is bigger than the dimension of the input signal, neurons form Gabor
wavelet-like receptive fields. For a smaller number of coding neurons, on
the other hand, the neurons develop center surround receptive fields,
pooling adjacent input dimensions.

\hypertarget{dendritic-balance-can-cope-with-long-transmission-delays}{%
\subsubsection*{Dendritic balance can cope with long transmission
delays}\label{dendritic-balance-can-cope-with-long-transmission-delays}}
\pdfbookmark[subsubsection]{Dendritic balance can cope with long transmission delays}{dendritic-balance-can-cope-with-long-transmission-delays}

A central problem for balanced networks is that long transmission delays
of inhibition can deteriorate network performance
\autocite{rullan_buxo_poisson_2019}. We found that DB networks are
much more robust to longer transmission delays than SB networks. To
investigate this, we simulated networks of 200 neurons with a range of time steps $\Delta t$, which we interpret as
transmission delays. We varied the delay from
$\Delta t = 0.1\text{ms}$ to $\Delta t =10\text{ms}$ and observed
how the delay affected coding performance for natural images.

\begin{figure}
\hypertarget{fig:delay}{%
\centering
\includegraphics[scale=0.5]{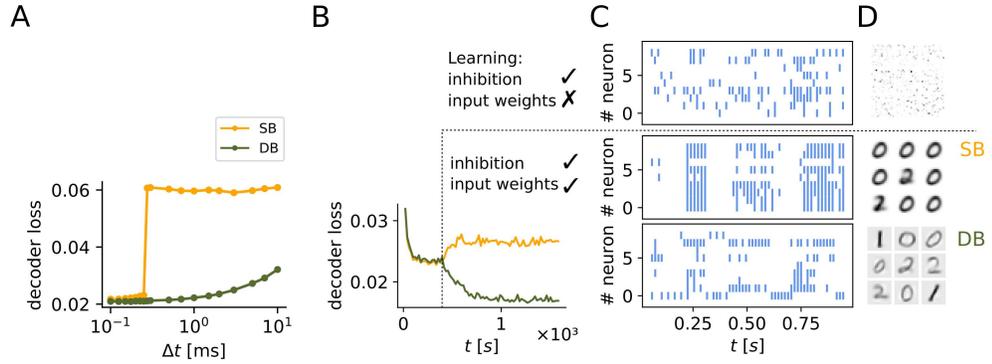}
\caption{\emph{Dendritic balance prevents learning of redundant representations for long transmission delays.} \textbf{A} Decoder loss of networks of 200 neurons coding for natural scenes for different transmission delays of inhibitory signals $\Delta t$. 
For transmission delays longer than $0.3 \, \text{ms}$, Hebbian-like learning (SB) leads to highly inefficient representations and large decoder loss. In contrast, for networks learning with dendritic balance (DB), the decoder loss increases only moderately even for long transmission delays. The results are robust with respect to the stochasticity of firing $\Delta u$ and the firing rate $\rho$ (see supplementary Figure \ref{fig:a7}).
\textbf{B-D} In another simulation experiment, networks of 9 neurons with long transmission delays of $3 \,\text{ms}$ received images of MNIST digits 0,1 and 2 as inputs. \textbf{B} To demonstrate the impact of lateral inhibitory and input plasticity, first only inhibition was learned; later plasticity on input weights was turned on (dashed line). Inhibitory plasticity decorrelates responses and decreases the decoder loss. When input plasticity was turned on, Hebbian-like plasticity (SB) learned worse representations than random feed-forward weights, which is indicated by the increase in decoder loss. In contrast, our model with dendritic balance (DB) learned improved representations with substantially reduced decoder loss. \textbf{C} The poor performance of the classical model (SB) is a consequence of highly synchronous spiking responses to the inputs, whereas neurons fire asynchronously in the model with dendritic balance (DB). \textbf{D} Neurons in the classical model (SB) learn overly similar feed-forward weights, whereas neurons with dendritic balance learn feed-forward weights that capture the input space well.}\label{fig:delay}
}
\end{figure}

We found that performance of SB networks drastically broke down to a
baseline level when transmission delays became longer than
$0.3\text{ms}$ (Fig~\ref{fig:delay}A). All neurons had learned the
same feed-forward weights (Fig~\ref{fig:a6}). In contrast, DB
networks continued to perform well even for much longer delays. While
long delays for DB also lead to a decrease in coding performance, DB
prevented the sudden collapse of the population code.

To illustrate the mechanism that caused the breakdown in performance for
SB, we also ran simulations of networks learning to code for MNIST
images with longer transmission delays (Fig~\ref{fig:delay}B). After
learning, the neurons endowed with Hebbian-like plasticity showed highly
synchronized activity (Fig~\ref{fig:delay}C) and had learned overly similar input weights (Fig~\ref{fig:delay}D). When
transmission delays become long, inhibition will often fail to prevent
that multiple neurons with similar input weights spike to encode the
same input. Hebbian-like plasticity can exacerbate this effect, since it
will adapt input weights of simultaneously spiking neurons in the same
direction, a vicious cycle which leads to highly pathological network
behaviour. In contrast, neurons learning with DB use the information
that inhibition provides for learning, even if it arrives too late to
prevent simultaneous spiking. Hence DB manages to learn distinct
features also in the face of long transmission delays.

\hypertarget{discussion}{%
\section*{Discussion}\label{discussion}}
\pdfbookmark[section]{Discussion}{discussion}

We asked how networks of spiking neurons can develop an efficient encoding of high-dimensional input signals with local plasticity rules.
Using a rigorous and top-down approach, we found that for the learning of efficient representations, single synapse plasticity has to take the population code into consideration.
Our results show that dendritic balance enables individual synapses to estimate the population code locally.
This can be harnessed by a voltage dependent learning rule that clearly outperforms Hebbian-like learning for naturalistic stimuli, or when inhibitory delays are present.

To learn efficient representations when neural responses are correlated, feed-forward plasticity has to incorporate the coding errors of the whole population.
Correlations between coding neurons in our model can arise through either correlations in the learned features or transmission delays of inhibitory feedback.
That learned features are correlated can in principle always be addressed by increasing the number of coding neurons (Fig~\ref{fig:scenes}), as this will increase the independence of learned features and hence reduce correlations between neurons.
Correlations due to transmission delays of inhibitory feedback, on the other hand, are a fundamental problem, which is independent of the precise architecture and known to occur in balanced networks \autocite{rullan_buxo_poisson_2019}.
In this case, Hebbian-like learning amplifies the correlations between neurons by adapting their feed-forward weights into the same direction, which ultimately can result in highly pathological network behaviour (Fig~\ref{fig:delay}).
Moreover, we show that learning by errors with dendritic balance overcomes the problem of transmission-delay induced correlations during learning (Fig~\ref{fig:delay}). 
We therefore argue that learning by errors becomes indispensable when transmission delays are present. 

In order to make coding errors available for single synapses locally, we introduced balanced dendritic potentials that are proportional to these errors.
Thereby, we extended the idea that coding errors can be presented by a tight balance at the soma \autocites{brendel2020learning}{deneve_efficient_2016} to a tight balance on individual dendrites. 
Since dendritic balance in our model also implies a somatic balance, it explains the same features of neural activity in the cortex: Highly irregular spiking \autocites{tolhurst1983statistical}{wohrer2013population}, but correlated membrane potentials of similarly tuned neurons \autocites{poulet2008internal}{yu2010membrane}.
Presenting an error through a balance of inputs is a general principle. By learning a balance through inhibitory plasticity, the network automatically learns an optimal decoding of neural activity to the excitatory inputs.
In principle it would therefore also be possible to present the coding error elsewhere, e.g. in the activity of other neural populations as suggested by predictive coding models \autocites{rao1999predictive}{bogacz2017tutorial}{friston2005theory}.
The advantage of presenting coding errors in local potentials, however, is that they are not rectified by neural spiking mechanisms, but instead are directly available as learning cues for synaptic plasticity.
What furthermore supports this idea is that a local balance of inputs, which is maintained by plasticity, has indeed been observed experimentally \autocites{liu_local_2004}{iascone2020whole}{bourne_coordination_2011}{hennequin2017inhibitory}.

In the presented model we assumed all-to-all inhibitory connectivity between coding neurons, which likely can be relaxed without losing the main benefits of the proposed learning scheme.
All-to-all inhibitory connections are problematic, since connecting every neuron to all dendrites of other neurons in the population would even for moderately sized networks prove extremely costly.
Moreover, in most neuron types the ratio of excitatory to inhibitory synapses is larger than one \autocite{villa2016excitatory}. 
However, not all of the inhibitory connections in our model are relevant. 
First, only inhibitory connections matter that really contribute in achieving a tight balance at the dendrites, i.e. when the dendritic potential is correlated to the presynaptic inhibiting neuron (Eq \ref{eq:dedriticbalance}).
Second, the number of inhibitory connections could be further decreased by clustering correlated excitatory synapses, which is a well known phenomenon \autocites{kastellakis_synaptic_2015}{chen_clustered_2012}{kleindienst_activity-dependent_2011}.
Clustered excitatory synapses could make use of the same inhibitory feedback if the inhibition provides a good estimate of the population code for all of these inputs. 
Third, inhibition is mediated by direct lateral connections between coding neurons, but less connections would be required if inhibition was mediated via interneurons. 
By incorporating inhibitory interneurons with broad feature selectivity, it would be possible to merge inhibitory connections that provide largely the same information. 
Finally, the number of inhibitory connections could be reduced by moving inhibition from the dendrites to the soma after learning. 
In our model, the second purpose of inhibition (besides informing excitatory plasticity) is to enable the concerted spiking of the neural population, which can also be guaranteed by somatic inhibition alone. 
Showing how exactly these reductions in wiring cost can be achieved, while maintaining the central benefits of the proposed learning scheme, should be a focus of future work.

A central element of our model is the dependence of synaptic plasticity on local membrane potentials, which has been argued to be a critical factor determining synaptic plasticity \autocites{lisman_postsynaptic_2005}{lisman2010questions}{artola1990different}{ngezahayo2000synaptic}.
Voltage-dependent plasticity is thought to be mediated mainly by the local Calcium concentration, which closely follows the local membrane potential \autocites{tsien_calcium_1990}{kanemoto2011spatial} and locally modulates neural plasticity \autocite{augustine2003local}. 
Classical voltage-dependent plasticity rules assume that synaptic plasticity happens only in a highly depolarized regime, where deviations from a critical potential determine the sign of the induced plasticity \autocites{clopath2010voltage}{shouval2002unified}.
Our derived rule can be reconciled with these classical rules if postsynaptic activity ($z_j$) brings the local potential into the highly depolarized regime, e.g. through back-propagating action potentials or dendritic plateau potentials \autocite{milojkovic2005strict}, and then local input ($u^i_j$) determines the sign of plasticity.
Furthermore, as required by our model, this voltage dependence implies that inhibition can have a major impact on excitatory synaptic plasticity locally \autocites{bar-ilan_role_2013}{saudargiene_inhibitory_2015}, which indeed has been found experimentally \autocites{wang2014inhibitory}{hennequin2017inhibitory}.
Therefore, while more work is necessary to understand what mechanisms might underlie its biophysical implementation, the proposed learning scheme is in principle compatible with voltage-dependent plasticity as mediated through local Calcium concentrations. 

Our model thus provides an answer to the major open question what the functional implications of plasticity mechanisms based on local dendritic potentials could be \autocites{hennequin2017inhibitory}{bono2017modelling}. 
Other studies have explored this issue (for a review see \autocite{bono2017modelling}), for example with a focus on feature binding \autocite{legenstein2011branch}. 
A theory related to ours suggests that voltage differences between soma and dendrite in two-compartment neurons present prediction errors \autocites{brea2016prospective}{guerguiev2017towards}{sacramento2018dendritic}. 
In contrast, our work shows that a coding error can be calculated from the mismatch between excitation and inhibition \textit{locally in individual dendritic compartments}, such that the local membrane potential represents the coding error.
We therefore propose that a central feature of synaptic plasticity based on local membrane potentials is to inform single synapse plasticity about plasticity-relevant activity of other neurons. 

Ultimately, we can generate two directly measurable experimental predictions from our model:
First, if input currents are locally unbalanced, inhibitory plasticity will learn to establish a local E-I balance.
Second, our model predicts that the strength of local inhibition determines the sign of synaptic plasticity:
During plasticity induction at excitatory feed-forward synapses, activating inhibitory neurons that target the same dendritic loci should lead to long-term depression of the excitatory synapses. 
We would expect this effect to persist, even if the inhibitory signal arrives shortly after the pre- and postsynaptic spiking. 

To conclude, we here presented a learning scheme that facilitates highly cooperative population codes for complex stimuli in neural populations. 
Our results question pairwise Hebbian learning as a paradigm for representation learning, and suggest that there exists a direct connection between dendritic balance and synaptic plasticity.

\hypertarget{Methods}{%
\section*{Methods}\label{methods}}
\pdfbookmark[section]{Methods}{Methods}

Neural activity was simulated in discrete timesteps of length $\Delta t$. Images were presented as continuous inputs for 100ms, after 70ms they were linearly interpolated to the next image in order to avoid discontinuities in the input signal. For every experiment a learning and a test set was created. The networks learned online on the training set; in regular intervals the learning rules were turned off and the performance was evaluated on the test set. Performance was measured via the instantaneous decoder loss (Eq \ref{eq:decoderloss}) by learning the decoder $D$ alongside the network. The respective update rule for the decoder is given by \begin{equation}\Delta D_{ij}\propto  z_j(x_i-\textstyle{\sum_k} D_{ik}z_k) .\label{eq:decodermethods}\end{equation}

For DB networks we propose two learning schemes with fast or slow inhibitory plasticity (detailed in SI). In order to reduce computation time for large networks, the analytical solution of optimal inhibitory weights $W^i_{jk} = - F_{ij}F_{ik}$ was used as an approximation of the proposed learning schemes. For Figure \ref{fig:mnist} the dendritic balance learning scheme with slow inhibitory plasticity is displayed. For the simulation of the correlated bars task (Fig~\ref{fig:bars}) and natural scenes (Fig~\ref{fig:scenes}), as well as Figure \ref{fig:delay} we used the analytical solution. When comparing the proposed learning schemes to the analytical solution on reference simulations (Fig~\ref{fig:a1}, \ref{fig:a2}) they consistently found very similar network parameters and reached the same performance.

In early simulations we observed that coding performance is largely affected by the population rate, i.e.~how many spikes can be used to encode the input signal. To avoid this effect when comparing the two learning schemes we additionally introduced a rapid compensatory mechanism to fix the firing rates which is realized by changing the thresholds $T_j$. We emphasize again that this adaptation is in principle not necessary to ensure stable network function. In fact, error-correcting balanced state inhibition can already be sufficient for a network to develop into a slow firing regime \autocite{brendel2020learning}. The fixed firing rate is enforced by adapting the threshold $T_j$ such that
neurons are firing with a target firing rate $\rho$.
\[\Delta T_j \propto (s_j -\rho \: \Delta t)\]
Here, $\rho \: \Delta t$ is the mean number of spikes in a time window
of size $\Delta t$ if a neuron would spike with rate $\rho$ and
$s_j$ is a spike indicator which is $1$ if $z_j$ spiked in the
last time $\Delta t$, otherwise $s_j=0$. 

We furthermore aided the learning process by starting with a high stochasticity in spiking and slowly decreasing it towards the desired stochasticity. We observed that convergence of the networks to an efficient solution often was faster and more reliable with this method. Specifically for the simulations of correlated bars and natural scenes we started with a stochasticity of $\Delta u = 1.0$. We then exponentially annealed it towards the final value $\Delta u^*$ by applying every timestep
\[\Delta u(t+1) = \Delta u(t) - \eta_{\Delta u} (\Delta u(t) - \Delta u^*).\]

Full derivations of the network dynamics and learning rules, supplementary figures containing additional information to simulation experiments, as well as simulation parameters, are provided in SI. Code for reproducing the main simulations is available online at \url{https://github.com/Priesemann-Group/dendritic_balance}.

\newpage

\pdfbookmark[section]{Acknowledgements}{Acknowledgements}
\hypertarget{Acknowledgements}{%
\section*{Acknowledgements}\label{Acknowledgements}}

We thank Friedemann Zenke and Christian Machens for their comments on the manuscript, as well as the Priesemann group, especially Matthias Loidolt and Daniel González Marx, for valuable comments and for reviewing the manuscript.
\textbf{Funding:} All authors received support from the Max-Planck-Society. L.R. acknowledges funding by SMARTSTART, the joint training program in computational neuroscience by the VolkswagenStiftung and the Bernstein Network. \textbf{Author contributions:} All authors designed the research. F.M. and L.R. conducted the research. F.M. implemented the simulations and created the figures. All authors wrote the paper. \textbf{Competing interests:} The authors do not declare any competing interests. \textbf{Data and materials availability:} Simulation and analysis code is available online at \url{https://github.com/Priesemann-Group/dendritic_balance}.
This work is licensed under a Creative Commons Attribution 4.0 International (CC BY 4.0) license, which permits unrestricted use, distribution, and reproduction in any medium, provided the original work is properly cited. To view a copy of this license, visit \url{https://creativecommons.org/licenses/by/4.0/}.

\phantomsection
\pdfbookmark[section]{References}{References}
\printbibliography

\newpage
\newcommand{\beginsupplement}{%
\setcounter{table}{0}
\renewcommand{\thetable}{S\arabic{table}}%
\setcounter{figure}{0}
\renewcommand{\thefigure}{S\arabic{figure}}%
}

\beginsupplement
\tableofcontents

\hypertarget{SI}{%
\section*{Supplementary Information}\label{SI}}
\pdfbookmark[section]{Supplementary Information}{SI}
\hypertarget{methodsglossary}{%
\subsection*{Symbols}\label{methodsglossary}}
\addcontentsline{toc}{subsection}{Symbols}

\small
\begin{multicols}{2}
\begin{itemize}
\item
  $\mathbf X_{0,T} = \{\mathbf{x}(t) | t\in \{0,...,T\}\}$: Input signal over time
  to be encoded
\item
  $\mathbf S_{0,T} = \{\mathbf{s}(t) | t\in \{0,...,T\}\}$: Spikes of
  coding neurons
\item
  $\mathbf{z}(t)$: `Outputs' of coding neurons, proportional to evoked
  post-synaptic potentials
\item
  $\mathbf{\hat x}(t)=D\mathbf{z}(t)$: Input signal reconstructed from network
  activity
\item
  $D$: Decoder matrix of the decoder model
\item
  $\sigma$: Variance of the decoder model
\item
  $\mathbf b$: Spiking probability prior of decoder model
\item
  $\theta$: Decoder model parameters $\{D,\sigma,\mathbf b\}$
\item
  $F$: Excitatory feed forward weights
\item
  $W$: Inhibitory recurrent weights connecting to the soma
\item
  $W^i$: Inhibitory recurrent weights, connecting to the dendrites to
  input $i$.
\item
  $T_j$: Soft threshold of neuron $j$
\item
  $\Delta u$: Stochasticity of neural spiking
\item
  $\tau$: Membrane time constant of leak
\item
  $\eta_{(\cdot)}$: learning rate of parameter $(\cdot)$
\item
  $\rho$: Target firing rate of neurons
\item
  $1/Z(\cdot)$: Normalization of probability function
\end{itemize}
\end{multicols}
\normalsize

\pagebreak

\hypertarget{stochastic-neural-dynamics}{%
\subsection*{Stochastic neural
dynamics}\label{stochastic-neural-dynamics}}
\addcontentsline{toc}{subsection}{Stochastic neural dynamics}

We simulated stochastic leaky integrate and fire neurons in discrete
timesteps. The model can be seen as a special case of the spike response model with escape noise \autocite{gerstner2002spiking}. In timestep $t\in\{0,1,...,T\}$ with length $\Delta t$
neuron $j$ spikes with a probability
\begin{equation}p_{\text{dyn}}\left(s_j(t)=1|\mathbf{x}(t),\mathbf{z}(t)\right)=p_{spike}(u_j(t))=\text{sig}\left(\frac{u_j(t) - T_j}{\Delta u}\right),\label{eq:pdyn}\end{equation}
where $\text{sig}(x)=[1+\exp(-x)]^{-1}$, $u_j(t)$ is the membrane
potential of the neuron, $T_j$ the firing threshold, $\Delta u$
defines how stochastic the spiking is and $s_j(t)$ is a spike
indicator, which is $1$ if neuron $j$ spiked in time step $t$,
otherwise $s_j=0$. Emitted spikes are then transmitted to other
neurons and elicit post synaptic potentials (PSPs) $\mathbf{z}(t)$
with
\[\begin{aligned} z_j(t) &= \sum_{t_s^j < t} \exp \left(-\frac{t-1-t_s^j}{\tau}\right),\end{aligned}\]
which account for the leaky integration at the membrane. Here, $t_s^j$
are the spike times of neuron $j$ and $\tau$ the membrane time
constant, which was chosen the same for all neurons. Please note that,
in order to ease the upcoming derivations, we changed notation such that
$t$ is the index of the discrete timestep and $\tau$ has the unit of
timesteps. The time delay of PSP arrival of the length of one time step
$\Delta t$ is interpreted as a finite traveling time of neural
impulses over axons. The EPSPs together with input signal
$\mathbf{x}(t)$ are then summed up linearly at the soma to give the
membrane potential
\[u_j(t)=\sum_i F_{ji} x_i(t)+\sum_{k} W_{jk}z_k(t).\]
In order to model neurons that make use of dendritic balance we
subdivided the somatic potentials such that they are sums of dendritic
potentials: $u_j(t) = \sum_i u^j_i(t)$, where the dendritic potentials
$u^j_i(t) = F_{ji}x_i(t) + \sum_k W^i_{jk} z_k(t)$. To summarize,
stochastic neural dynamics are modeled through the spike probability
$p_{\text{dyn}}\left(\mathbf{s}(t)|\mathbf{x}(t),\mathbf{z}(t)\right)$
with neural parameters $\{F,W,\mathbf T,\Delta u \}$.

\hypertarget{learning-an-efficient-code-with-expectation-maximization}{%
\subsection*{Learning an efficient code with expectation
maximization}\label{learning-an-efficient-code-with-expectation-maximization}}
\addcontentsline{toc}{subsection}{Learning an efficient code with expectation maximization}

With the following derivations we provide a link between learned balanced state inhibition \autocite{brendel2020learning} and neural sampling in graphical models \autocite{nessler_bayesian_2013}. Furthermore we will address the linear case of the quite general problem which arises through explaining away effects, i.e.~converging arrows in graphical models: Converging arrows imply that neurons should cooperate to encode the input and lead to non-localities in update rules when the neural dynamics are based on point neurons. In related studies this problem so far has been avoided in various ways, which all prevent the network from explaining the input through possibly correlated neurons simultaneously and thus limit coding versatility \autocites{nessler_bayesian_2013}{bill_distributed_2015}{nessler2009stdp}{kappel_stdp_2014}{deneve_bayesian_2007}.

The goal of neural spiking dynamics and plasticity throughout this paper
is to find an efficient spike encoding, i.e.~representing an input
signal $\mathbf{X}_{0,T} = \{\mathbf x(t) | t\in \{0,...,T\} \}$
through a collection of spikes
$\mathbf{S}_{0,T} = \{\mathbf s(t) | t\in \{0,...,T\} \}$.
$\mathbf{X}_{0,T}$ can be seen here as an episode in an organisms
life, which we will assume to be distributed according to
$p^*(\mathbf{X}_{0,T})$. We say that $\mathbf{S}_{0,T}$ efficiently
encodes $\mathbf{X}_{0,T}$ if the following two conditions are met:
\begin{enumerate}
\def\labelenumi{\alph{enumi})}
\item
  $\mathbf{X}_{0,T}$ can be accurately estimated from
  $\mathbf{S}_{0,T}$ via a decoding model
  $p_\theta(\mathbf{X}_{0,T}|\mathbf{S}_{0,T})$.
\item
  The number of spikes emitted is small.
\end{enumerate}
Hence we want to maximize the likelihood
$p_\theta(\mathbf{X}_{0,T}|\mathbf{S}_{0,T})$ over both the decoding
model parameters $\theta$ and the latent variables
$\mathbf{S}_{0,T}$ sampled by the (constrained) network dynamics
$p_{\text{dyn}}\left(\mathbf{s}(t)|\mathbf{x}(t),\mathbf{z}(t)\right)$.

\begin{figure}
\hypertarget{fig:graphicalmodel}{%
\centering
\includegraphics[scale=0.7]{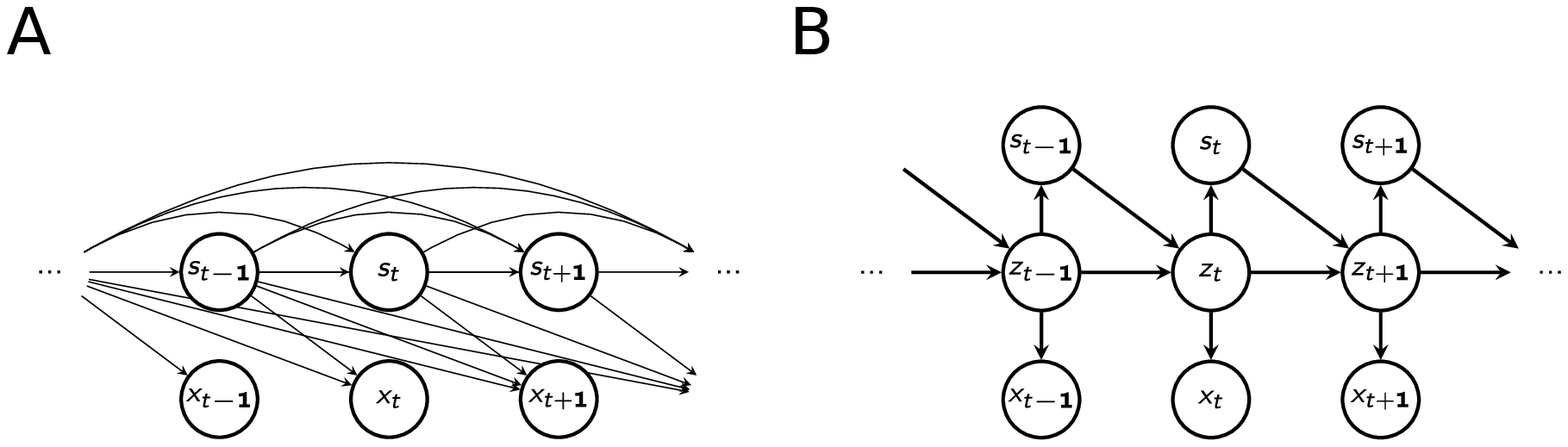}
\caption{Graphical representation of the decoder model. \textbf{A} We
consider a decoding model where readouts of inputs $\mathbf x(t)$
(denoted here as $\mathbf x_t$) are conditioned on preceding spikes
$\mathbf s(t)$ (denoted as $\mathbf s_t$). \textbf{B} By introducing the spike traces
$\mathbf z(t)$ into the model, the model factorizes over timesteps,
which is equivalent to viewing it as a hidden Markov model (HMM) with
hidden states
$\{\mathbf z(t),\mathbf s(t)\}$.}\label{fig:graphicalmodel}
}
\end{figure}

To show how a stochastic spiking neural network can unsupervisedly learn
such an encoding, we make use of the framework of online
expectation-maximization (EM) learning \autocite{neal1998view}.
EM-learning can find maximum-likelihood estimates for parameters of
latent variable models (here
$p_\theta(\mathbf{X}_{0,T},\mathbf{S}_{0,T})$) for observed data
($\mathbf{X}_{0,T}$). For these models it typically is intractable to
marginalize out the latent variables ($\mathbf{S}_{0,T}$). In order to
solve this problem one defines the log-likelihood lower bound
\begin{equation}\begin{aligned} \mathcal{F}^*(\tilde p, \theta) &= 
\left\langle\log p_\theta(\mathbf{X}_{0,T})- D_{KL}(\tilde p(\mathbf{S}_{0,T}|\mathbf{X}_{0,T})| p_\theta(\mathbf{S}_{0,T}|\mathbf{X}_{0,T})) \right\rangle_{p^*(\mathbf{X}_{0,T})} \\ 
&= \langle \log p_\theta(\mathbf{X}_{0,T},\mathbf{S}_{0,T}) - \log \tilde p(\mathbf{S}_{0,T}|\mathbf{X}_{0,T})\rangle_{\tilde p(\mathbf{S}_{0,T}|\mathbf{X}_{0,T})p^*(\mathbf{X}_{0,T})}\end{aligned},\label{eq:lllowerbound}\end{equation}
where $\tilde p(\mathbf{S}_{0,T}|\mathbf{X}_{0,T})$ can be any
(tractable) probability distribution, which in our case will be given
through $p_{dyn}$. Finding maximum-likelihood parameters $\theta$
can then be done by iteratively maximizing
$\mathcal{F}^*(\tilde p, \theta)$ with respect to $\tilde p$
(E-step) and $\theta$ (M-step). In the E-step $p_\theta$ is
approximated by $\tilde p$ in order to estimate
$\langle \log p_\theta(\mathbf{X}_{0,T},\mathbf{S}_{0,T})\rangle_{\tilde p(\mathbf{S}_{0,T}|\mathbf{X}_{0,T})p^*(\mathbf{X}_{0,T})}$
and in the M-step this approximation is used to improve the model. This
algorithm is guaranteed to converge to a local minimum, also if
$\mathcal{F}^*$ is maximized only partially in every iteration, which
makes it applicable to online learning.

Appealing to this theory in the following we show that: (i) Given a
linear decoding model, a stochastic spiking neural network can be
connected such that it can sample an efficient encoding online. This
relates model- and network-parameters. (ii) The decoding model can be
optimized online in respect to the sampled dynamics of the network. (iii)
Combining (i) (the E-step) and (ii) (the M-step) yields update rules that
can be applied by a stochastic spiking neural network to optimize its
parameters in order to encode its inputs. 

\hypertarget{online-encoding-by-spiking-neural-network}{%
\subsubsection*{Online encoding by spiking neural
network}\label{online-encoding-by-spiking-neural-network}}
\addcontentsline{toc}{subsubsection}{Online encoding by spiking neural
network}

Let us consider the following decoding model and prior on the spiking
probability (Fig~\ref{fig:graphicalmodel})
\[p_\theta(\mathbf{X}_{0,T}|\mathbf{S}_{0,T}) = \prod_t p_\theta(\mathbf{x}(t)|\mathbf{z}(t)) = \prod_t \mathcal{N}_{\mathbf x(t)}(D\mathbf z(t), \Sigma)\]
\[p_{\theta}(\mathbf{S}_{0,T}) = \prod_t p_{\theta}(\mathbf{s}(t)|\mathbf{z}(t)) = \prod_t \frac{1}{Z(\mathbf{b})}\exp\left(\mathbf s(t)^\top \mathbf b \right)\]
with $\Sigma=\sigma^2 I$ and parameters
$\theta = \{D,\sigma,\mathbf b\}$. Notably this model asserts that at
every time $t$, $\mathbf x(t)$ can be linearly decoded from spike
traces $\mathbf z(t)$ with variance $\sigma^2$, where the spike
traces are defined as before. Observe that the spike traces
$\mathbf z(t)$ are deterministically defined given the preceding
spikes $\mathbf{S}_{0,t-1}$. Since the model factorizes over time
given the spike traces $\mathbf z(t)$, the log-likelihood lower bound
(Eq \ref{eq:lllowerbound}) can be rewritten as
\[\begin{aligned}\mathcal{F}^*(p_{dyn}, \theta) =& \left\langle \sum_t \log p_\theta(\mathbf{x}(t),\mathbf{s}(t)|\mathbf{z}(t)) - \log p_{dyn}(\mathbf{s}(t)|\mathbf{x}(t),\mathbf{z}(t))\right\rangle_{p_{dyn}(\mathbf{S}_{0,T}|\mathbf{X}_{0,T})p^*(\mathbf{X}_{0,T})} \\
=& \left\langle \log p_\theta(\mathbf{X}_{0,T})\right\rangle_{p^*(\mathbf{X}_{0,T})} - \\
& \left\langle \sum_t \log p_{dyn}(\mathbf{s}(t)|\mathbf{x}(t),\mathbf{z}(t)) - \log p_\theta(\mathbf{s}(t)|\mathbf{X}_{t+1,T},\mathbf{z}(t))\right\rangle_{p_{dyn}(\mathbf{S}_{0,T}|\mathbf{X}_{0,T})p^*(\mathbf{X}_{0,T})} 
\end{aligned}\]

Here we substituted
$\tilde p(\mathbf{S}_{0,T}|\mathbf{X}_{0,T})=\prod_t p_{dyn}(\mathbf{s}(t)|\mathbf{x}(t),\mathbf{z}(t))$
and made use of the fact that spikes alter only the future decoding
since they are independent of the past given $\mathbf z(t)$, i.e.
$p_\theta(\mathbf{s}(t)|\mathbf{X}_{0,T},\mathbf{S}_{0,t-1}) = p_\theta(\mathbf{s}(t)|\mathbf{X}_{t+1,T},\mathbf{z}(t))$.

We now perform the E-step. $\mathcal{F}^*$ is approximately maximized
over $p_{\text{dyn}}$ if
$p_{\text{dyn}}(\mathbf{s}(t)|\mathbf{x}(t),\mathbf{z}(t))\approx p_\theta(\mathbf{s}(t)|\mathbf{X}_{t+1,T},\mathbf{z}(t))$
at every time $t$. However, this poses two problems: (i)
$p_{\text{dyn}}$ depends only on $\mathbf x(t)$ while the spike
probability in the model is based on future values
$\mathbf{X}_{t+1,T}$, which are not available to the network. (ii) In
order to compute
$p_\theta(\mathbf{s}(t)|\mathbf{X}_{t+1,T},\mathbf{z}(t))=\sum_{\mathbf{S}_{t+1,T}} p_\theta(\mathbf{S}_{t,T}|\mathbf{X}_{t+1,T},\mathbf{z}(t))$
future spikes have to be marginalized out, which is intractable. For the
purpose of this paper we introduced simple approximations that solve
these problems and work well in practice for our inputs. Specifically we
assumed input- and network activity to be approximately constant over
time. Hence all future inputs $\mathbf x(t')\in\mathbf{X}_{t+1,T}$
were assumed to be known to be $\mathbf x(t')=\mathbf x(t)$. Future
network activity (independent of the current spike $\mathbf s(t)$) was
assumed to be well approximated by a single trajectory, where neural
outputs $\mathbf z(t)$ were constant. With this we can compute

\begin{align*}\sum_{\mathbf{S}_{t+1,T}} & p_\theta(\mathbf{s}(t)|\mathbf{X}_{t+1,T},\mathbf{z}(t),\mathbf{S}_{t+1,T})p_\theta(\mathbf{S}_{t+1,T}|\mathbf{X}_{t+1,T},\mathbf{z}(t))\\
\approx&\prod_{t'=t}^T p_\theta\left(\mathbf{s}(t)|\mathbf x(t')=\mathbf x(t),\mathbf{z}(t')=\mathbf{z}(t)+\mathbf s(t) \exp\left(-\frac{t'-1-t}{\tau} \right)\right)\\
=& \frac{1}{Z(\theta, \mathbf x)} \exp\left( \mathbf s(t)^\top \mathbf b \right)\prod_{t'=t+1}^T \exp\left( \frac{\mathbf z(t')^\top}{\sigma^2}[D^\top\mathbf x(t') - \frac{1}{2}D^\top D \mathbf z(t')] \right)\\
=& \frac{1}{Z(\theta, \mathbf x)} \exp\left( \mathbf s(t)^\top \mathbf b \right)\prod_{t'=t+1}^T \exp\Bigg( \frac{\left(\mathbf{z}(t)+\mathbf s(t) \exp\left(-\frac{t'-1-t}{\tau} \right)\right)^\top}{\sigma^2}  \Bigg[D^\top\mathbf x(t) - \\ 
&- \frac{1}{2}D^\top D \left(\mathbf{z}(t)+\mathbf s(t) \exp\left(-\frac{t'-1-t}{\tau} \right)\right)\Bigg] \Bigg)\\
=& \frac{1}{Z(\theta, \mathbf x, \mathbf z)} \exp\left( \mathbf s(t)^\top \left[\mathbf b + \frac{\tau}{\sigma^2}D^\top\mathbf x(t) - \frac{\tau}{\sigma^2}D^\top D \left(\mathbf{z}(t)+\frac{1}{4}\mathbf s(t)\right) \right]\right)\\
=& \frac{1}{Z(\theta, \mathbf x, \mathbf z)} \exp\left( \mathbf s(t)^\top \left[\mathbf b + \frac{\tau}{\sigma^2}D^\top\mathbf x(t) - \frac{\tau}{\sigma^2}D^\top D \mathbf{z}(t)-\frac{1}{4}\frac{\tau}{\sigma^2}\text{diag} (D^\top D) \right]\right)\end{align*}
where we approximated
$\sum^T_{t'=t+1}\exp (-\frac{t'-1-t}{\tau})\approx\tau$ (that is
$\tau$ and $T$ large) and the last equality follows if timesteps are
sufficiently small such that only one neuron spikes per timestep.
Comparing with the network dynamics (Eq \ref{eq:pdyn}) from this we can
conclude that a network that performs approximate online sampling from
$p_\theta(\mathbf{S}_{0,T}|\mathbf{X}_{0,T})$ has parameters
\begin{equation}\begin{aligned}
F&=D^\top \\
W &= - D^\top D \\
T_j &= \frac{1}{4} W_{jj} - \frac{\sigma^2}{\tau}b_j \\
\Delta u&= \frac{\sigma^2}{\tau}
\end{aligned}\label{eq:networkparams}\end{equation}
These results are similar to those yielded by a greedy spike encoding
scheme \autocite{brendel2020learning}. 
Please note that the sampling could be improved by using
advanced sampling schemes, such as rejection sampling
\autocite{kappel_stdp_2014}.

\hypertarget{online-learning-of-an-optimal-decoder}{%
\subsubsection*{Online-learning of an optimal
decoder}\label{online-learning-of-an-optimal-decoder}}
\addcontentsline{toc}{subsubsection}{Online-learning of an optimal
decoder}

As we have shown, the network dynamics implement an approximation of the
E-step if the network parameters are chosen correctly. We will now use
these samples produced by the network to incrementally improve the
parameters of the decoding model in the M-step.

Recall that in the M-step we want to maximize under $\theta$
\[\left\langle \sum_t \log p_\theta(\mathbf{x}(t),\mathbf{s}(t)|\mathbf{z}(t))\right\rangle_{p_{dyn}(\mathbf{S}_{0,T}|\mathbf{X}_{0,T})p^*(\mathbf{X}_{0,T})}.
\]
Updates of the decoder model parameters should thus follow the gradient
\[\Delta \theta = \tilde \eta_\theta \frac{\partial \mathcal{F}^*}{\partial \theta}= \tilde \eta_\theta  \left\langle \sum_t \frac{\partial}{\partial \theta} \log p_\theta(\mathbf{x}(t),\mathbf{s}(t)|\mathbf{z}(t))\right\rangle_{p_{dyn}(\mathbf{S}_{0,T}|\mathbf{X}_{0,T})p^*(\mathbf{X}_{0,T})}\]
In this paper we're only interested in the decoder weights $D_{ij}$
from neuron $j$ to input $i$, where the derivation yields
\[\Delta D_{ij} = \tilde \eta_D  \left\langle \sum_t \sigma^{-2}z_j(t)\left(x_i(t) - \sum_k D_{ik} z_k(t) \right) \right\rangle_{p_{dyn}(\mathbf{S}_{0,T}|\mathbf{X}_{0,T})p^*(\mathbf{X}_{0,T})}\]
Here, $\tilde{\eta}_D$ is the update step size and $\sigma^2$ the
variance of the decoder model. Note that in the following we will drop
the dependence of the learning rate on $\sigma^2$, which has its
motivation in covariant optimization \autocite{mackay2003information}.
In covariant optimization, the gradient is multiplied by the inverse
curvature of the loss function, because step size should be decreased
when the curvature of the loss function is high. Since the curvature of
the likelihood is proportional to the inverse variance, the variance
drops out to yield a covariant gradient. This yields the update rule
\[\Delta D_{ij} = \tilde \eta_D  \left\langle \sum_t z_j(t)\left(x_i(t) - \sum_k D_{ik} z_k(t) \right) \right\rangle_{p_{dyn}(\mathbf{S}_{0,T}|\mathbf{X}_{0,T})p^*(\mathbf{X}_{0,T})}\]

\textbf{Online approximation} If many episodes $\mathbf{X}_{0,T}$ as sampled from
$p^*(\mathbf{X}_{0,T})$ are presented in succession and spikes are
sampled as outlined above, the average over samples from $p_{dyn}$ can
be replaced by an average over time
\[\left\langle \sum_t \cdot \right\rangle_{p_{dyn}(\mathbf{S}_{0,T}|\mathbf{X}_{0,T})p^*(\mathbf{X}_{0,T})}\approx \sum_t \left\langle \cdot \right\rangle_t.\]

If the update rules are performed every timestep this lets us rewrite
them as
\begin{equation}\delta D_{ij} = \eta_D z_j(t)\left(x_i(t) - \sum_k D_{ik} z_k(t) \right)\label{eq:decoderupdate}\end{equation}
This requires, however, that the learning rate $\eta_D$ is
sufficiently small such that changes in $D_{ij}$ are negligible in a
sufficiently long time window $T'$. In that case, summing the equation
over time window $T'$ yields
\[\begin{aligned} \sum_{t=0}^{T'}\delta{D_{ij}} &=\eta_D T' \left\langle z_j(t)\left(x_i(t) - \sum_k D_{ik} z_k(t) \right) \right\rangle_{t=0}^{T'}\\
&\overset{T'\rightarrow T}{\approx} \Delta D_{ij},
\end{aligned}\]
where the learning rates are related via $\tilde{\eta}_D=\eta_D T'$. A
more refined statement can be made by rewriting the update equation as
\[\Delta D_{ij} = \tilde \eta_D \left( \left\langle z_j(t) x_i(t) \right\rangle_{p_{dyn}(\mathbf{S}_{0,T}|\mathbf{X}_{0,T})p^*(\mathbf{X}_{0,T})} - \sum_k D_{ik} \left\langle z_j(t) z_k(t) \right\rangle_{p_{dyn}(\mathbf{S}_{0,T}|\mathbf{X}_{0,T})p^*(\mathbf{X}_{0,T})}\right)\]
This makes explicit that the only information required to compute the
gradient of the decoder weights are the correlations between neural
outputs and inputs and in between neural outputs over the input sequences.
Thus in practice, the learning rate $\eta_D$ is ideally chosen as
large as possible to allow fast learning, but also sufficiently small
such that weight updates are performed with respect to a time window
that provides a good estimate of correlations under the whole sampled
spike trains.

\hypertarget{online-learning-of-network-parameters}{%
\subsubsection*{Online-learning of network
parameters}\label{online-learning-of-network-parameters}}
\addcontentsline{toc}{subsubsection}{Online-learning of network
parameters}

So far we showed that the parameters of a network that samples from a
decoder model are directly connected to the parameters of the model. We
also showed how the decoder weights have to be updated such that they
maximize the model likelihood over the generated samples. We will now
combine these two results to find update rules for neural parameters
directly, that can be used by neurons to learn an efficient encoding
without supervision online. To this end we will first show how an
approximation to the previously derived gradients can be implemented by
regular stochastic LIF neurons. In a second step we will show how a
better approximation can be realized by neurons with dendritic
compartments. The central insight for all derivations will be that
learning an E-I balance on membrane potentials
corresponds to the learning of a decoder to the excitatory inputs times
a transformation matrix that brings them into the space of the membrane
potentials.

\textbf{Somatic balance
approximation}

\textit{Feed-forward weights} \\*[1ex]
From the equality $F=D^\top$ (Eq
\ref{eq:networkparams}) derived earlier and the update rule for $D$
(Eq \ref{eq:decoderupdate}) we directly arive at
\[\delta F_{ji} = \eta_F  z_j(t)\left(x_i(t) - \sum_k F_{ki} z_k(t) \right) \]
We follow previous approaches \autocite{brendel2020learning} and ommit
contributions to the decoding $\sum_k F_{ki} z_k(t)$ that are not
available for the neuron, which is equivalent to assuming that neural
spiking in the population is uncorrelated
$\forall j\neq k: \langle z_j(t) z_k(t) \rangle_t \approx 0$. This
yields the regularized Hebbian rule

\begin{equation}\delta F_{ji} = \eta_F  z_j(t)\left(x_i(t) - F_{ji} z_j(t)\right) \label{eq:methodshebbianlearning}\end{equation}

\textit{Inhibitory weights} \\*[1ex]
This rule will follow the optimal decoder
gradient if spikes are indeed uncorrelated. However, if this is not the
case the solution will be suboptimal and furthermore the previously
derived inhibition $W=-D^\top D$ together with the suboptimal weights
$F$ does not enable a reasonable encoding anymore. Both problems can
be addressed by observing that for the optimal membrane potential we
derived
\[\mathbf u^{opt}(t) = D^\top \mathbf x(t) - D^\top D \mathbf z(t) = D^\top (\mathbf x(t) - D \mathbf z(t)) \]
i.e.~the potentials are proportional to the decoding error. This can be
approximately guaranteed even if the feed-forward weights are suboptimal
(but not zero) by setting $W=-FD$, since then
\[\mathbf u(t) = F \mathbf x(t) - F D \mathbf z(t) = F (\mathbf x(t) - D \mathbf z(t)) \overset{\approx}{\propto} \mathbf u^{opt}(t)\]

To make sure that neurons adapt their encoding for an optimal decoder,
inhibitory weights will adapt along the gradient of decoder weights. For
fixed encoder weights $F$ this yields
\begin{equation}\begin{aligned}
\delta {W}_{jk} &= - \sum_i F_{ji} \delta {D}_{ik} \\
&= -\eta_W z_k(t)\left(\sum_i F_{ji}x_i(t) - \sum_{i,l} F_{ji}D_{il} z_l(t)\right) \\
&= -\eta_W z_k(t)\left(\sum_i F_{ji}x_i(t)  + \sum_{l} W_{jl} z_l(t)\right)  \\
&= -\eta_W z_k(t) u_j(t)
\end{aligned}\label{eq:methodssomaticbalance}\end{equation}
This shows that through an E-I balance, this rule for $W$
self-consistently finds the correct decoder `inside' of the inhibitory
weights, and hence allows the projection of the right decoding error
$x-Dz$ through lateral inhibition. Thereby the inhibition ensures a
reasonable encoding even if feed-forward weights are not learned
optimally. Since in the equation above $F_{ji}$ is assumed constant,
we chose the learning rate $\eta_W$ 2-4 times larger than $\eta_F$.
In simulations we found that inhibitory weights that evolve under Eq
\ref{eq:methodssomaticbalance} indeed converged to $W= - FD$, where
$D$ is the optimal decoder weights obtained under the non-local update
rule Eq \ref{eq:decoderupdate} .

\textbf{Learning encoder weights with dendritic
balance
} In the following we devise examples for local plasticity rules for
feed-forward inputs that follow the correct gradient of the likelihood
lower bound. Locality requires that the decoding of other neurons is
made available at the synapse, which can then be used to find the
correct gradient. We argue that this can be mediated by dendritic
inhibitory connections $W^i$ that target dendrites where the
feed-forward input $i$ has formed a synapse. Due to the locality in
space, the membrane potential $u^i_j$ in the vicinity of synapse $i$
on that dendrite only integrates inputs that are present locally, i.e.
\begin{equation}\begin{aligned}
u^i_j(t) = \underbrace{F_{ji} x_i(t)}_{\text{excitatory input}} + \underbrace{\textstyle{\sum_{k}} W^i_{jk} z_k(t)}_{\text{dendritic inhibition}} .
\end{aligned}\label{eq:dendriticpotentialmethods}\end{equation}
We then assume a regime where currents from the dendrites are summed
linearly, such that the total membrane potential at the soma is given by
$u_j(t)= \sum_i u_j^i(t)$. 
Similar to somatic inhibition, we will show that
dendritic inhibitory connections can locally learn an optimal decoding
of neural PSPs $\mathbf{z}$ by enforcing \emph{dendritic balance} of
excitatory and inhibitory inputs. The central feature of this approach
is that inhibition and excitation both use the dendritic potential for
learning, which requires their cooperation. We here show two possible
mechanisms that realize this and yield very similar behaviour to the
analytical solution (Fig~\ref{fig:a1},
\ref{fig:a2}).

\textit{Slow feed-forward adaptation} \\*[1ex]
One possibility to ensure the cooperation of feed-forward and inhibitory
weights is to separate the timescales on which they are adapting. To
that end we make the optimal inhibition ansatz similar to before
$W_{jk}^i= -F_{ji} D_{ik}$. Then, changing inhibitory weights in the
direction of the decoder gradient of Eq \ref{eq:decoderupdate} yields
\[\begin{aligned}
\delta {W}_{jk}^i =& - F_{ji} \delta {D}_{ik} \\
=& -\eta_W z_k(t) (F_{ji} x_i(t) - \textstyle{\sum_l} F_{ji}D_{il}z_l(t))\\
=& -\eta_Wz_k(t) (F_{ji} x_i(t) + \textstyle{\sum_l} W_{jl}^iz_l(t))\\
=& -\eta_Wz_k(t) u_j^i(t).
\end{aligned}\]
where we again assumed that changes in feed-forward weights are slow and
can be neglected, and $\eta_W=\eta_D$. We conclude that enforcing
dendritic balance by inhibitory plasticity is equivalent to locally
optimizing a decoder $D_{ik} = - W_{jk}^i/F_{ji}$.

The correct gradient of the decoder weights can also be calculated
locally, but it can't be applied to the feed-forward weights directly
since this would contradict the previously made assumption of slow
changes in feed-forward weights. However, it is possible to locally
integrate the correct gradient and use this to adapt feed-forward
weights slowly, with a delay. To this end we introduce the local
integration variable $I_{ji} = F_{ji}D_{ij}$, which adapts according
to the decoder gradient times $F_{ji}$
\[\begin{aligned}
 \delta {I}_{ji} =&  \eta_{I} F_{ji} z_j(t) \left( x_i(t) - \textstyle{\sum_{k}} D_{jk}z_k(t) \right)\\
 =&  \eta_{I} z_j(t) u^i_j(t)\\ 
 \end{aligned}\] with $\eta_{I}=\eta_D$. $F_{ji}$ then can slowly
follow $D_{ij}$ via
\[\begin{aligned}
\delta {F}_{ji} =&  \eta_{F} (I_{ji}/F_{ji} - F_{ji})\\ 
\end{aligned}\]
with $\eta_F \ll \eta_D$. Note that for $F_{ji}=0$ the gradient for
$F_{ji}$ is not defined. In this case the learning process could be
kickstarted via simple Hebbian learning on $F_{ji}$. It is also
possible to assume very small but nonzero weights from the start. Note also that the equation $W_{jk}^i = -F_{ji}D_{ik}$ has to hold at the start of learning, which can be guaranteed by simply choosing $W_{jk}^i = F_{ji} = 0$. To
summarize, slow feed-forward adaptation leads to neural parameters
$W^i_{jk} = -F_{ji} D_{ik}$ and $F_{ji}=D_{ij}$. This shows that
input synapses slowly can evolve to minimize the decoder error along its
gradient using only local information.

\textit{Simultaneous adaptation of feed-forward and inhibitory weights} \\*[1ex]
In principle it would also be possible to adapt feed-forward and
inhibitory weights simultaneously without a separation of timescales.
However, calculating the gradient for the derived inhibitory weights is
locally not feasible, since we find
\[\begin{aligned}
\delta {W}_{jk}^i =& - D_{ji} \delta {D}_{ik} - \delta D_{ji} {D}_{ik} \\
=& -\eta_D(z_k(t) u_j^i(t) + z_j(t) u_k^i(t)).
\end{aligned}\]
Empirically we found that the contralateral contributions
$z_j(t) u_k^i(t)$ to the gradient can be approximated by the
accessible contributions $z_k(t) u_j^i(t)$. We thus approximate
$\langle z_j(t) u_k^i(t)\rangle_t \approx \langle z_k(t) u_j^i(t)\rangle_t$.
While this equation does not hold for all $i,j,k$, we still find that
the resulting learned contributions to the dendritic potentials have the
correct magnitude, hence enabling feed-forward learning. The gradient
for the inhibitory weights now are
\[\begin{aligned}
\delta {W}_{jk}^i=& -\eta_W z_k(t) u_j^i(t),
\end{aligned}\]
where $\eta_W=2\eta_D$.

Assuming the correct inhibition ${W}_{jk}^i=-D_{ji}D_{ki}$ we can find
the decoded population encoding locally at the dendrite. From the
self-consistency $F_{ji}=D_{ij}$ and Eq
\ref{eq:dendriticpotentialmethods} we have the relation
\[\sum_{k} D_{jk}z_k(t) = \frac{F_{ji} x_i(t) - u^i_j(t)}{F_{ji}}\]

With this we can implement the learning of feed-forward weights in way
that highlights its similarity to previous approaches (Eq
\ref{eq:methodshebbianlearning})
\[\delta F_{ji} = \eta_F  z_j(t)\left(x_i(t) - \frac{F_{ji} x_i(t) - u^i_j(t)}{F_{ji}} \right), \]
i.e.~the rule is a regularized Hebbian plasticity rule. Again, for very
small $F_{ji}$ the regularization becomes unstable, but can be left
away (since it should go to zero) leaving a purely Hebbian rule. For the
derivation we used $\eta_F=\eta_D$, which implies that we should
choose $\eta_W\approx 2\eta_F$. A relatively slow inhibitory learning
that coincides with feed-forward plasticity has also been reported by
\autocite{froemke2007synaptic}.

In simulations we verified that the approximations we made for this
learning scheme are adequate and yield feed forward weights for which
$F_{ji}=D_{ij}$ holds with high accuracy. Note that the network found
by the presented learning scheme only corresponds to the decoding model
if $\eta_W\approx 2\eta_F$. However, if the inhibitory learning is
faster this only results in a rescaling of feed-forward weights by a
factor of $2\eta_F / \eta_W$, since their adaptation is too slow by
this factor. This means that in this case the ``correct'' dynamics of
the network can be recovered via a rescaling of all weights, or equivalently, with firing rate adaptation a
change in the stochasticity of spiking $\Delta u$ by a factor of
$2\eta_F / \eta_W$.

\textbf{Rapid firing rate adaptation} In the Bayesian framework Habenschuss and colleagues have 
shown that a rapid rate adaptation can be interpreted as a
constraint on the variational approximation in the E-step
\autocite{habenschuss2012homeostatic}. For the resulting constrained
optimization formally a Lagrange multiplier is introduced which
`overwrites' the analytic threshold $T_j=\frac{1}{4}W_{jj} - \sigma^2\tau^{-1} b_j$. We
will not make a notational difference between the two thresholds here. The fixed firing rate is enforced by adapting the threshold $T_j$ such that
neurons are firing with a target firing rate $\rho$.
\[\delta T_j = \eta_T (s_j -\rho \: \Delta t)\]
Here, $\rho \: \Delta t$ is the mean number of spikes in a time window
of size $\Delta t$ if a neuron would spike with rate $\rho$ and
$s_j$ is a spike indicator which is $1$ if $z_j$ spiked in the
last time $\Delta t$, otherwise $s_j=0$. Since this is a constraint
that is applied in the E-step, the learning rate $\eta_T$ should be
large.

\hypertarget{datasets}{%
\subsection*{Datasets}\label{datasets}}
\addcontentsline{toc}{subsection}{Datasets}

\textbf{MNIST}

The standard MNIST database of handwritten digits was used
\autocite{lecun2010mnist}. Images were scaled down from $28\times 28$
to $16\times 16$ pixels. No further preprocessing was applied.

\textbf{Correlated bars}

See description in Figure \ref{fig:bars}A. Pixels where bars are
displayed (also in the case of overlap) were set to the value 1.0,
pixels without bars were set to 0.0.

\textbf{Natural scenes}

Images for natural scenes were taken from \autocite{bruno}. A simple
preprocessing was applied to ensure that they can be modeled by spiking
neurons. Importantly we required that input is always positive. Every
image $\mathbf{\chi}$ in the database was whitened, which can be seen
as an approximation of retinal on/off-cell preprocessing, where one
on-cell and one off-cell with overlapping fields are lumped together in
a single value $\chi_i$ which can be positive or negative. We
separated every value $\chi_i$ into two values $x'_{2i}=\chi_i$ and
$x'_{2i+1}=-\chi_i$. We then applied a continuous nonlinear activation
function to ensure that activations are positive and bimodally distributed (i.e. mostly close to either 0.0 or 1.0): $x_i = \text{sig}(3.2 (x'_i - 0.8) )$, where $\text{sig}(x)=1/(1+\exp{(-x)})$. For the display of
learned weights we merge corresponding values again and display
$x_{2i} - x_{2i+1}$.
\newpage
\hypertarget{parameters}{%
\subsection*{Parameters}\label{parameters}}
\addcontentsline{toc}{subsection}{Parameters}

For all tasks parameters were tuned to ensure that networks operate
well. \textit{DB} denotes networks where the analytic solution given the decoder
was used for network dynamics. \textit{DB delayed} are networks with slow
feed-forward adaptation, \textit{DB simultaneous} are networks with parallel
adaptation of feed-forward and inhibitory weights. \textit{SB} are networks
learning with somatic balance. When the parameter $\eta_{\Delta u}$ is
present the stochasticity of spiking was annealed starting from 1.0 with
rate $\eta_{\Delta u}$.

\textbf{MNIST} (Fig~\ref{fig:mnist}, \ref{fig:delay}, \ref{fig:a1})

\begin{tabular}{lllll} \hline\hline
Parameter & DB & DB simultaneous & DB delayed & SB\\ \hline
$\Delta t$ & 0.1ms & 0.1ms & 0.1ms & 0.1ms \\
$\tau$ & 10ms& 10ms& 10ms& 10ms \\
$\Delta u$ & 0.1 & 0.1 & 0.1 & 0.1 \\
$\rho$ & 20$\text{s}^{-1}$ & 20$\text{s}^{-1}$ & 20$\text{s}^{-1}$ & 20$\text{s}^{-1}$ \\
$\eta_T$ & $5.0\cdot10^{-3}$  & $3.0\cdot10^{-3}$  & $5.0\cdot10^{-4}$  & $5.0\cdot10^{-3}$ \\
$\eta_F$ & -                  & $3.0\cdot10^{-6}$ & $4.0\cdot10^{-7}$ & $5.0\cdot10^{-6}$ \\
$\eta_I$ & -                  & -                 & $4.0\cdot10^{-5}$ & - \\
$\eta_W$ & -                  & $6.0\cdot10^{-6}$  & $4.0\cdot10^{-5}$  & $1.0\cdot10^{-5}$ \\
$\eta_D$ & $5.0\cdot10^{-6}$  & $3.0\cdot10^{-6}$  & $5.0\cdot10^{-6}$  & $5.0\cdot10^{-6}$ \\\hline\hline
\end{tabular}

\textbf{Correlated bars} (Fig~\ref{fig:bars}, \ref{fig:a2})

\begin{tabular}{lllll} \hline\hline
Parameter & DB & DB simultaneous & DB delayed & SB\\ \hline
$\Delta t$ & 1.0ms & 1.0ms & 1.0ms & 1.0ms \\
$\tau$ & 10ms& 10ms& 10ms& 10ms \\
$\Delta u^*$ & 0.1 & 0.1 & 0.1 & 0.1 \\
$\eta_{\Delta u}$ & $7.0\cdot10^{-8}$ & $7.0\cdot10^{-8}$ & $7.0\cdot10^{-8}$ & $7.0\cdot10^{-8}$ \\
$\rho$ & 15$\text{s}^{-1}$ & 15$\text{s}^{-1}$ & 15$\text{s}^{-1}$ & 15$\text{s}^{-1}$ \\
$\eta_T$ & $1.0\cdot10^{-2}$  & $1.0\cdot10^{-2}$  & $5.0\cdot10^{-2}$  & $1.0\cdot10^{-2}$ \\
$\eta_F$ & -                  & $5.0\cdot10^{-5}$ & $1.0\cdot10^{-7}$ & $5.0\cdot10^{-5}$ \\
$\eta_I$ & -                  & -                 & $5.0\cdot10^{-5}$ & - \\
$\eta_W$ & -                  & $1.0\cdot10^{-4}$  & $5.0\cdot10^{-5}$  & $1.0\cdot10^{-4}$ \\
$\eta_D$ & $5.0\cdot10^{-5}$  & $5.0\cdot10^{-5}$  & $5.0\cdot10^{-5}$  & $5.0\cdot10^{-5}$ \\\hline\hline
\end{tabular}

\begin{figure}
\addcontentsline{toc}{subsubsection}{\textbf{Fig \ref{fig:a3}} All learning curves for the natural scenes task.}
\hypertarget{fig:a3}{%
\centering
\includegraphics[scale=0.6]{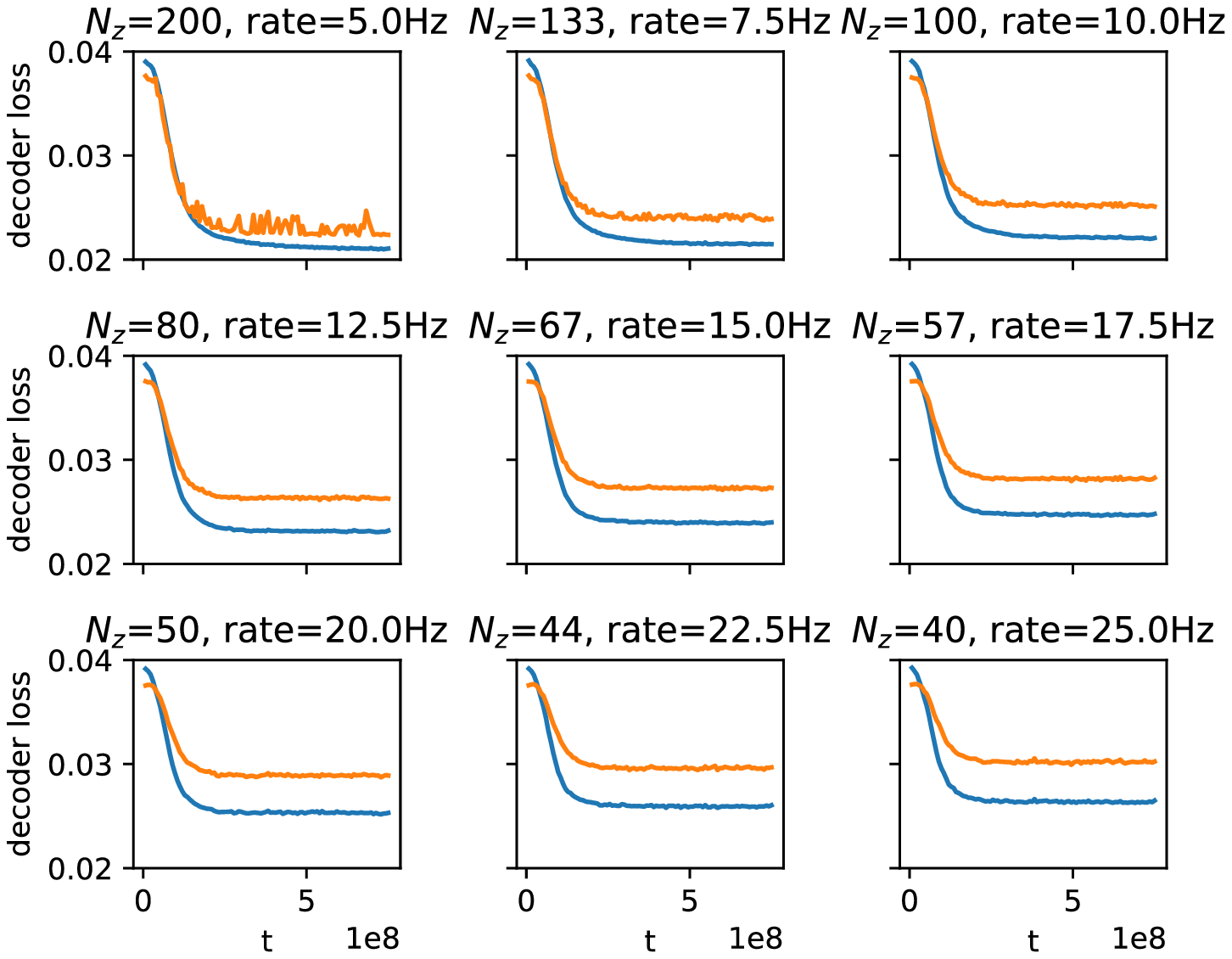}
\caption{All learning curves for the natural scenes task (Fig
\ref{fig:scenes}).}\label{fig:a3}
}
\end{figure}

\begin{figure}
\addcontentsline{toc}{subsubsection}{\textbf{Fig \ref{fig:a4}} All learned weights for the natural scenes task.}
\hypertarget{fig:a4}{%
\centering
\includegraphics[scale=0.5]{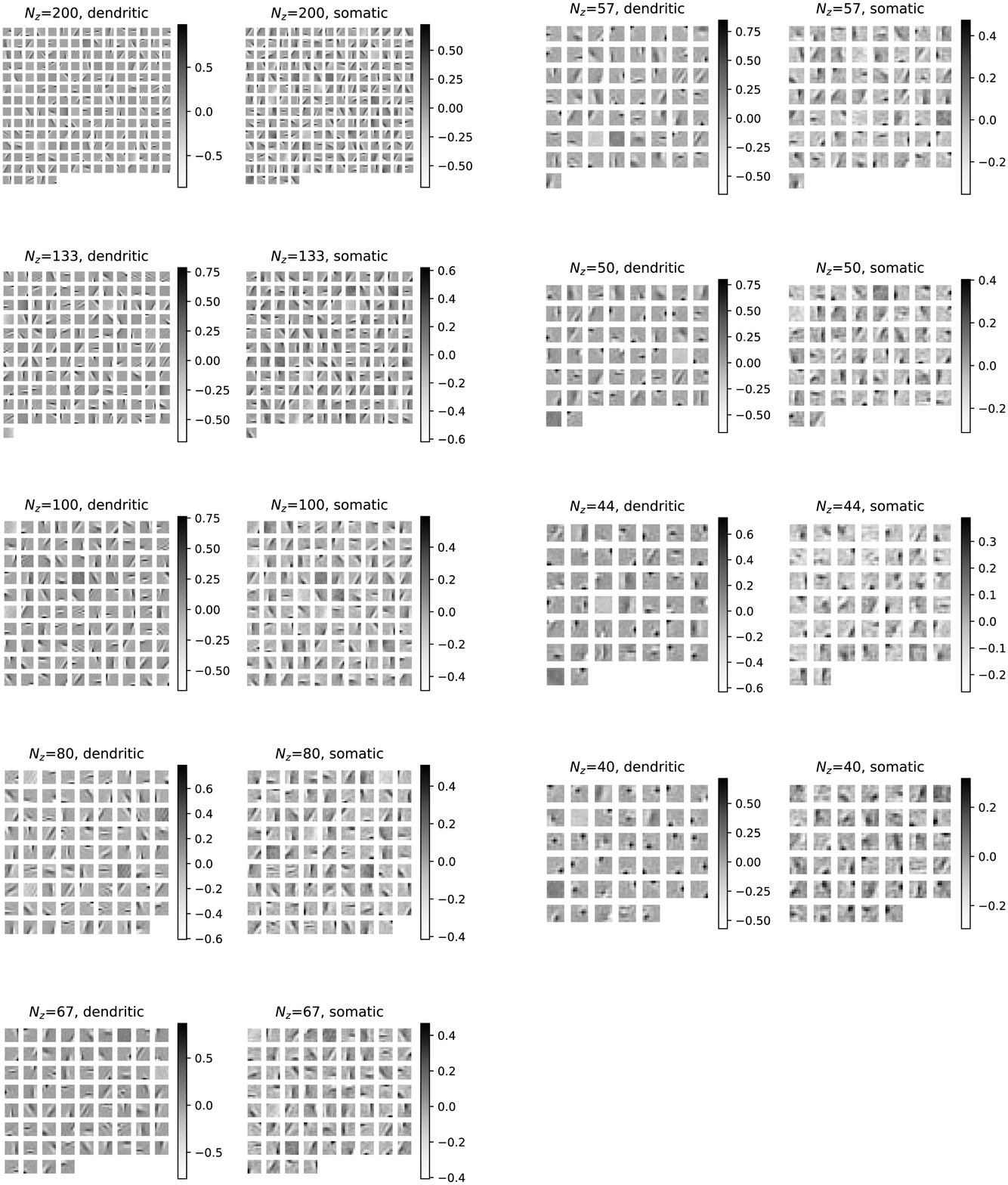}
\caption{All learned weights for the natural scenes task (Fig
\ref{fig:scenes}).}\label{fig:a4}
}
\end{figure}

\begin{figure}
\addcontentsline{toc}{subsubsection}{\textbf{Fig \ref{fig:a5}} All learning curves for the natural scenes task for different timesteps.}
\hypertarget{fig:a5}{%
\centering
\includegraphics[scale=0.6]{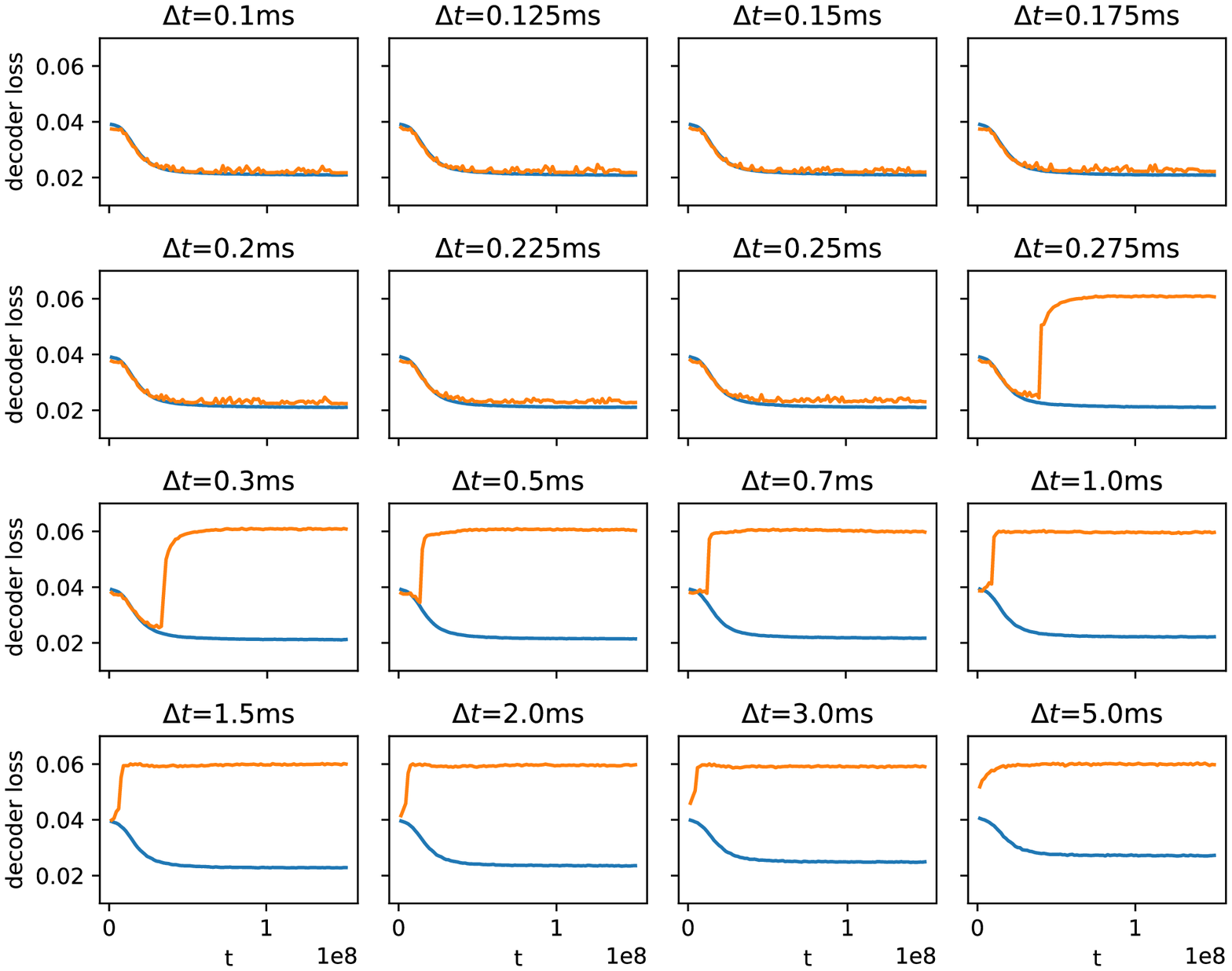}
\caption{All learning curves for the natural scenes task for different
timesteps (Fig~\ref{fig:delay}A).}\label{fig:a5}
}
\end{figure}

\begin{figure}
\addcontentsline{toc}{subsubsection}{\textbf{Fig \ref{fig:a7}} The results in Figure \ref{fig:delay}A are robust in respect to the stochasticity of firing $\Delta u$ and firing rate $\rho$.}
\hypertarget{fig:a7}{%
\centering
\includegraphics[scale=0.7]{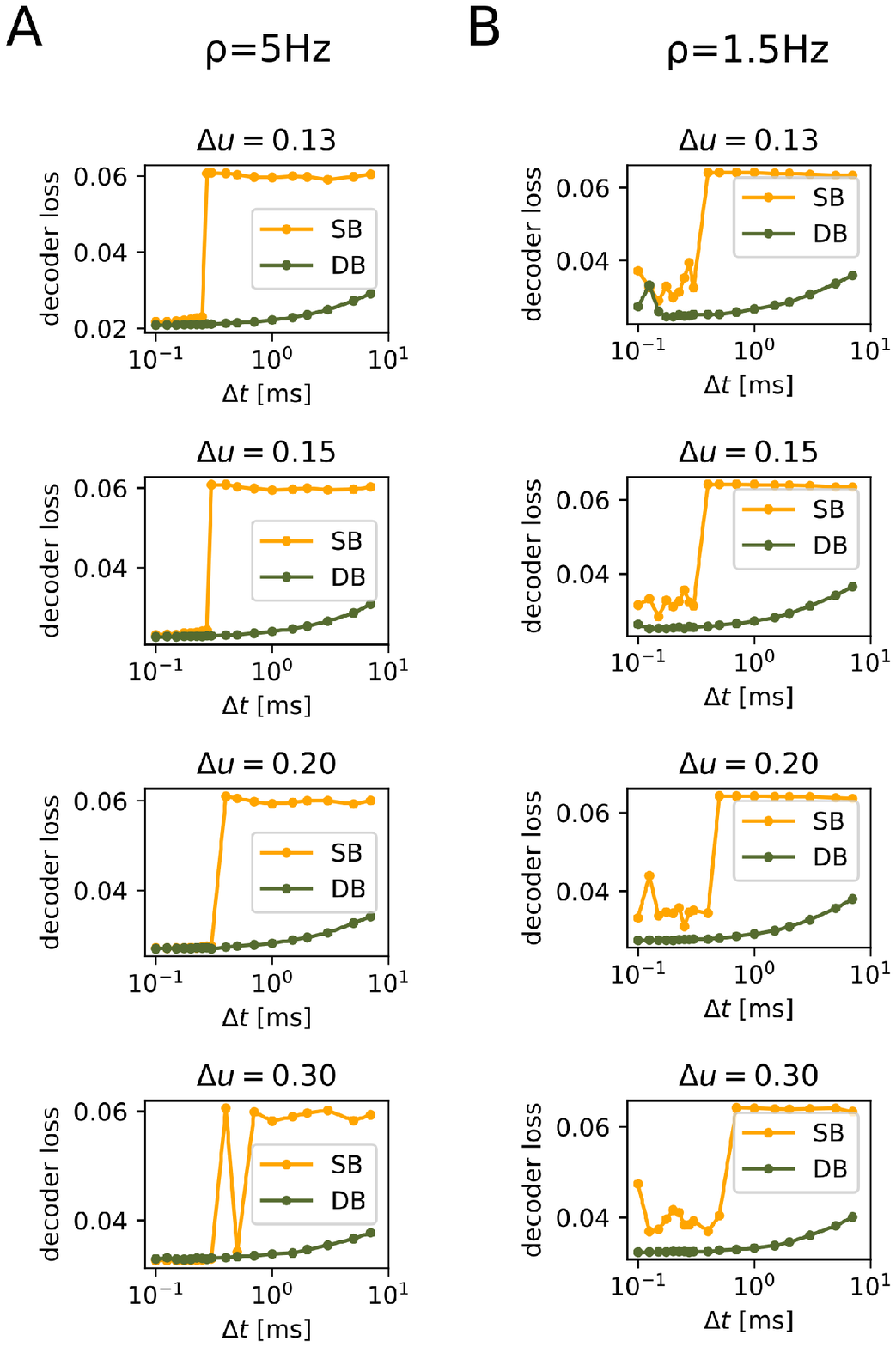}
\caption{The results in Figure \ref{fig:delay}A are robust in respect to the stochasticity of firing $\Delta u$ and firing rate $\rho$. We tested firing rates of \textbf{A} $\rho=5\text{Hz}$, where learning is mostly stable, and \textbf{B} $\rho=1.5\text{Hz}$, where learning becomes slightly unstable. For higher stochasticity (larger $\Delta u$) neural firing becomes extremely random, for more deterministic neurons (smaller $\Delta u$) learning often does not converge.}\label{fig:a7}
}
\end{figure}

\textbf{Natural scenes} (Fig~\ref{fig:scenes}, \ref{fig:delay}, 
\ref{fig:a3}, \ref{fig:a4}, \ref{fig:a5}, \ref{fig:a6}, \ref{fig:a7})

\begin{tabular}{lll} \hline\hline
Parameter & DB & SB\\ \hline
$\Delta t$ & 0.2ms & 0.2ms \\
$\tau$ & 10ms  & 10ms \\
$\Delta u^*$ & 0.13  & 0.13 \\
$\eta_{\Delta u}$ & $7.0\cdot10^{-8} $  & $7.0\cdot10^{-8}$  \\
$\rho \cdot \text{\# neurons}$ & 1000$\text{s}^{-1}$ & 1000$\text{s}^{-1}$\\
$\eta_T$ until $t=6000$s & $6.0\cdot10^{-3}$ & $6.0\cdot10^{-3}$ \\
$\eta_T$ until $t=\infty$ & $4.0\cdot10^{-3}$ & $4.0\cdot10^{-3}$ \\
$\eta_F$ until $t=6000$s & - & $4.0\cdot10^{-5}$ \\
$\eta_F$ until $t=\infty$ & - & $3.0\cdot10^{-5}$ \\
$\eta_W$ until $t=6000$s & - & $10.0\cdot10^{-5}$ \\
$\eta_W$ until $t=\infty$ & - & $7.0\cdot10^{-5}$ \\
$\eta_D$ until $t=6000$s & $4.0\cdot10^{-5}$ & $4.0\cdot10^{-5}$ \\
$\eta_D$ until $t=\infty$ & $3.0\cdot10^{-5}$ & $3.0\cdot10^{-5}$ \\\hline\hline
\end{tabular}

\pagebreak

\addcontentsline{toc}{subsection}{Supplementary figures}
\hypertarget{supplementary-figures}{%
\subsection*{Supplementary figures}\label{supplementary-figures}} 
\begin{figure}[!ht]
\addcontentsline{toc}{subsubsection}{\textbf{Fig \ref{fig:a1}} Comparison of the different learning schemes on the MNIST task.}
\hypertarget{fig:a1}{%
\centering
\includegraphics[scale=0.7]{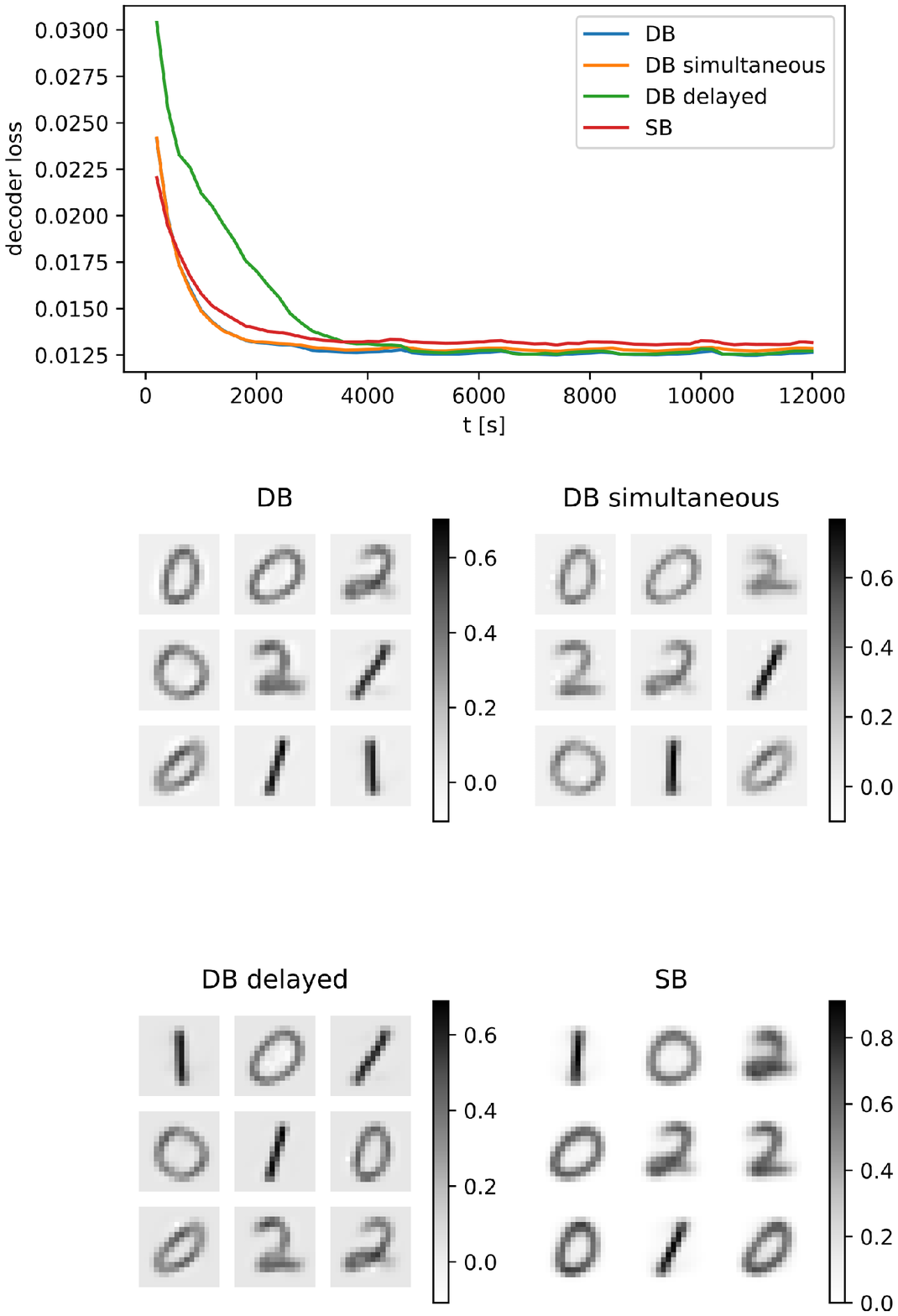}
\caption{Comparison of the different learning schemes on the MNIST
task.}\label{fig:a1}
}
\end{figure}

\begin{figure}
\addcontentsline{toc}{subsubsection}{\textbf{Fig \ref{fig:a2}} Comparison of the different learning schemes on the bars task with $p=0.8$.}
\hypertarget{fig:a2}{%
\centering
\includegraphics[scale=0.7]{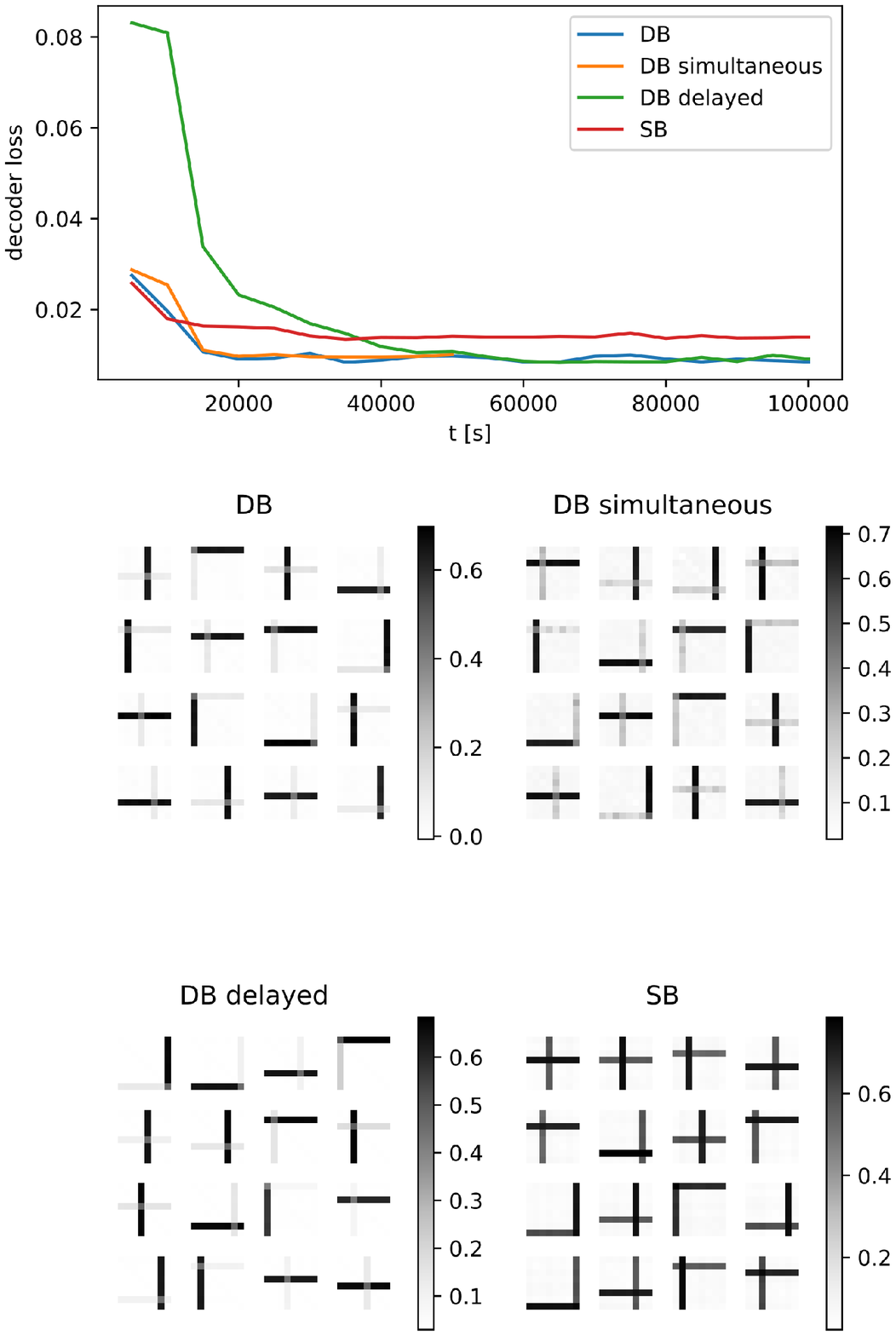}
\caption{Comparison of the different learning schemes on the bars task
with $p=0.8$.}\label{fig:a2}
}
\end{figure}

\begin{figure}
\addcontentsline{toc}{subsubsection}{\textbf{Fig \ref{fig:a6}} All learned weights for the natural scenes task for different timesteps.}
\hypertarget{fig:a6}{%
\centering
\includegraphics[scale=0.3]{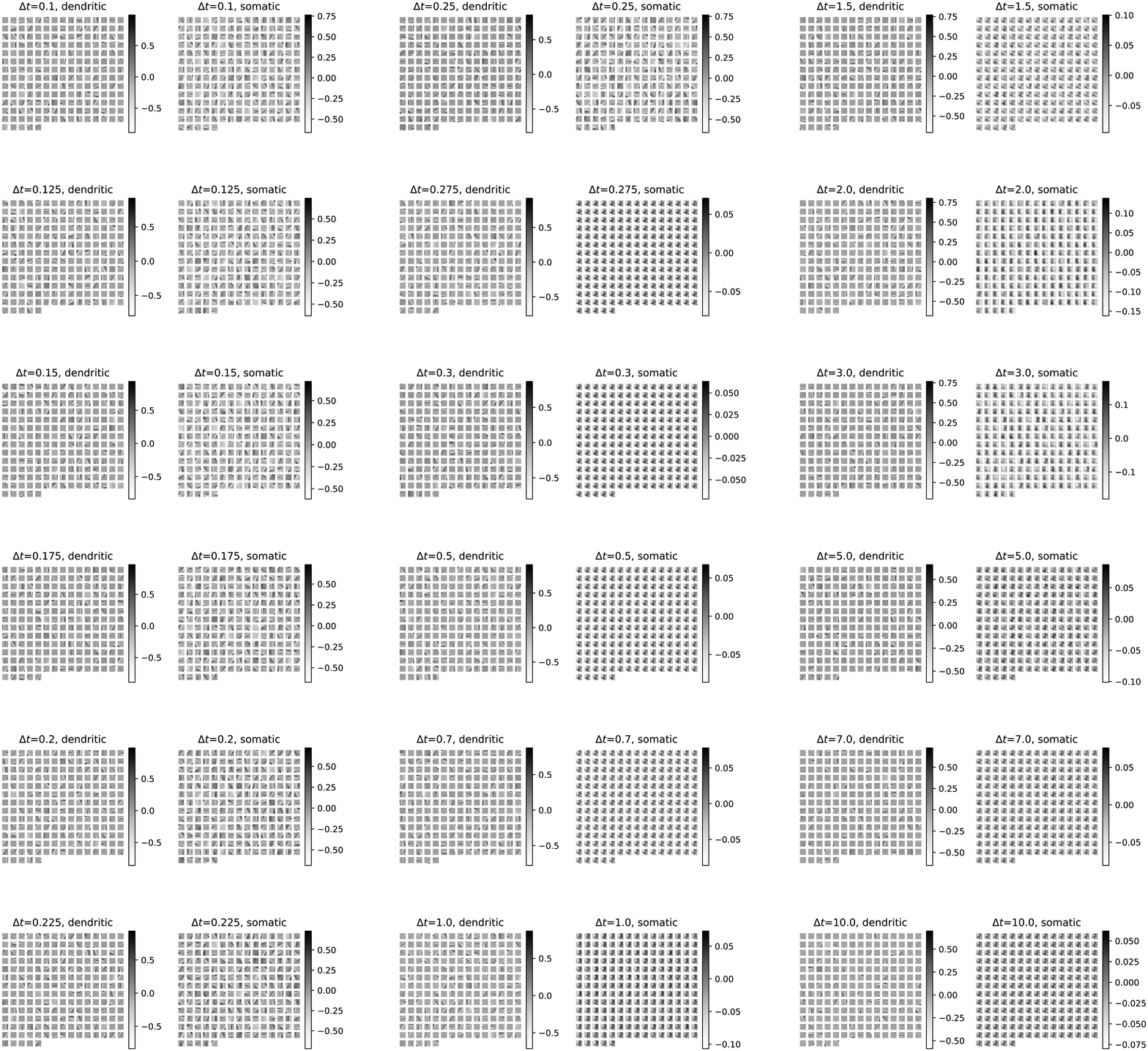}
\caption{All learned weights for the natural scenes task for different
timesteps (Fig~\ref{fig:delay}A).}\label{fig:a6}
}
\end{figure}

\end{document}